\documentclass[11pt,a4]{article}
\usepackage{amssymb,amsmath,graphicx}
\usepackage{amsfonts,setspace}
\usepackage[caption=false]{subfig}
\usepackage{epsfig,latexsym,graphicx,color,url}
\newtheorem{theorem}{Theorem}[section]

\newtheorem{remark}{Remark}[section]
\newtheorem{definition}{Definition}[section]

\begin{document}

\title{Consistent Fixed-Effects Selection in Ultra-high dimensional Linear Mixed Models with Error-Covariate Endogeneity}

\author{Abhik Ghosh$^1$ and Magne Thoresen$^2$\\
$^1$ Indian Statistical Institute, India\\
$^2$ Department of Biostatsitics, University of Oslo, Norway}
\date{}
\maketitle

\begin{abstract}
Recently, applied sciences, including longitudinal and clustered studies in biomedicine 
require the analysis of ultra-high dimensional linear mixed effects models 
where we need to select important fixed effect variables from a vast pool of available candidates.
However, all existing literature assume that all the available covariates and random effect components are independent
of the model error which is often violated (endogeneity) in practice. 
In this paper, we first investigate this important issue 
in ultra-high dimensional linear mixed effects models with particular focus 
on the fixed effects selection. We study the effects of different types of endogeneity
on existing regularization methods and prove their inconsistencies.  
Then, we propose a new profiled focused generalized method of moments (PFGMM)
approach to consistently select fixed effects under `error-covariate' endogeneity, i.e., 
in the presence of correlation between the model error and covariates.
Our proposal is proved to be oracle consistent with probability tending to one 
and works well under most other type  of endogeneity too.
Additionally, we also propose and illustrate a few consistent parameter estimators,
including those of the variance components, along with variable selection through PFGMM. 
Empirical simulations and an interesting  real data example further support the claimed utility of our proposal.
\end{abstract}


\section{Introduction}\label{SEC:intro}


Linear mixed effects models are widely used for analysis of clustered data in econometrics, 
biomedicine and other applied sciences. 
It consists of additional random-effect components, along with the usual fixed-effects regression modeling, 
to account for variability among clusters. 
In biomedical applications, typical examples are longitudinal studies with repeated measurements within individuals 
and multi-center studies with patients clustered within centers. 
Due to recent technological advances, we often have access to sets of extremely high-dimensional explanatory variables, 
typically so-called omics data, in such studies. 
Hence, the potential fixed effects variables are often in the order of millions even in studies with relatively few patients. 
Thus, we have to select the important fixed-effects variables from the vast pool of available variables under an ultra-high dimensional set-up. 
Note that, in most such studies the relevant random effect variables are typically few, 
and  their selection is not necessary.

Mathematically, given $I$ groups (e.g., centers) indexed by $i=1,2, \ldots, I$,
we observe $n_i$ responses in the $i$-th group, denoted by the $n_i$-dimensional vector $\boldsymbol{y}_i$.
The associated fixed and random effect covariate values are, respectively  denoted by  the $n_i\times p$ matrix $\boldsymbol{X}_i$
and the $n_i\times q$ matrix $\boldsymbol{Z}_i$; often $\boldsymbol{Z}_i$ is a subset of $\boldsymbol{X}_i$. 
Let $n=\sum_{i=1}^I n_i$ denote the total number of observations. 
Then, the linear mixed model (LMM) is defined as (Pinheiro and Bates, 2000)

\begin{equation}
\boldsymbol{y}_i=\boldsymbol{X}_i\boldsymbol{\beta} + \boldsymbol{Z}_i\boldsymbol{b}_i + \boldsymbol{\epsilon}_i,~~~~~i=1,...,I,
\label{EQ:model1}
\end{equation}
where $\boldsymbol{\beta}$ is the fixed effects (regression) coefficient vector, $\boldsymbol{b}_i$s are the random effects 
and $\boldsymbol{\epsilon}_i$s are random error components in the model. 
We assume that $\boldsymbol{b}_i \sim N_q(0, \boldsymbol{\Psi}_{\boldsymbol{\theta}})$ and  
$\boldsymbol{\epsilon}_i\sim N_{n_i}(0, \sigma^2\boldsymbol{I}_{n_i})$, for each $i=1, \ldots, I$, 
and they are independent of each other and also of the $\boldsymbol{X}_i$s.
Here $\boldsymbol{I}_d$ denotes the identity matrix of order $d$ and 
$\boldsymbol{\Psi}_{\boldsymbol{\theta}}$ is a model variance matrix defined in terms of a $q^*$-dimensional (unknown) 
parameter vector $\boldsymbol{\theta}$; e.g., $\boldsymbol{\Psi}_{\boldsymbol{\theta}} = Diag\{\theta_1, \ldots, \theta_{q} \}$ with $q^*=q$,
or $\boldsymbol{\Psi}_{\boldsymbol{\theta}}=\theta_1 \boldsymbol{I}_{q}$ with $q^*=1$, etc. 
Then, given $\boldsymbol{X}_i$ (and $\boldsymbol{Z}_i$), 
$\boldsymbol{y}_i\sim N_{n_i}(\boldsymbol{X}_i\boldsymbol{\beta}, \sigma^2 \boldsymbol{V}_i(\boldsymbol{\theta},\sigma^2))$,
independently for each $i$, where $\boldsymbol{V}_i(\boldsymbol{\theta},\sigma^2)
=\sigma^{-2}\boldsymbol{Z}_i\boldsymbol{\Psi}_{\boldsymbol{\theta}}\boldsymbol{Z}_i^T + \boldsymbol{I}_{n_i}$.
Stacking the variables in larger matrices, we can rewrite the LMM (\ref{EQ:model1}) as

\begin{equation}
\boldsymbol{y}=\boldsymbol{X}\boldsymbol{\beta} + \boldsymbol{Z}\boldsymbol{b} + \boldsymbol{\epsilon},
\label{EQ:model0}
\end{equation}
%
where $\boldsymbol{y}=(\boldsymbol{y}_1^T, \ldots, \boldsymbol{y}_I^T)^T$, 
$\boldsymbol{X}=(\boldsymbol{X}_1^T, \ldots, \boldsymbol{X}_I^T)^T$,
$\boldsymbol{Z}={\rm Diag}\{\boldsymbol{Z}_1, \ldots, \boldsymbol{Z}_I\}$,
$\boldsymbol{b}=(\boldsymbol{b}_1^T, \ldots, \boldsymbol{b}_I^T)^T
\sim N_{qI}(\boldsymbol{0}, \boldsymbol{I}_{q}\otimes\boldsymbol{\Psi}_{\boldsymbol{\theta}})$
and $\boldsymbol{\epsilon}=(\boldsymbol{\epsilon}_1^T, \ldots, \boldsymbol{\epsilon}_I^T)^T\sim N_{n}(\boldsymbol{0}, \sigma^2\boldsymbol{I}_{n})$.
Now given $\boldsymbol{X}$, 
$\boldsymbol{y}\sim N_{n}(\boldsymbol{X}\boldsymbol{\beta}, \sigma^2 \boldsymbol{V}(\boldsymbol{\theta},\sigma^2))$
with $\boldsymbol{V}={\rm Diag}\{\boldsymbol{V}_1, \ldots, \boldsymbol{V}_I\}$.
We wish to perform inference about the unknown regression parameter $\boldsymbol{\beta}$
and the variance parameter vector $\boldsymbol{\eta}=(\boldsymbol{\theta}, \sigma^2)$.
As described in the beginning, in this paper, we assume that $p\gg n$, but 
$q\ll p, n$, and prefixed.
In particular, we assume the ultra-high dimensional set-up with $\log p = O(n^\alpha)$ for some $\alpha\in (0,1)$,
which is often the case with omics data analysis. 
Then the total number of parameters is $d:=p+q^\ast+1 \gg n$ 
and we need to impose a sparsity condition for estimating $\boldsymbol{\beta}$.
This entitles a selection of important fixed-effect variables;
we assume that the number of such variables is $s\ll n$.
Then, the selection of these $s$ important fixed-effect variables 
and the parameter estimation are done by maximizing a suitably penalized log-likelihood function 
(Schelldorfer, Buhlmann and Van de Geer, 2011;  Ghosh and Thoresen, 2018; Fan and Li, 2012); 
the resulting estimator of $(\boldsymbol{\beta},\boldsymbol{\eta})$ is known as 
the maximum penalized likelihood estimator (MPLE); see Section \ref{SEC:MPLE}.

Under our ultra-high dimensional regime, an important desired property of the MPLE is the oracle variable selection consistency 
which ensures that all the true important variables and only those are selected 
with probability tending to one, asymptotically. 
All the existing literature study this property of the MPLE under the crucial assumption of independence 
of the covariates $\boldsymbol{X}$ with the model error $\boldsymbol{\epsilon}$ and the random effects $\boldsymbol{b}$;
these independence assumptions are referred to as the `\textit{exogeneity}' of the model. 
However, they may not always hold in practice, and the corresponding situation is referred to as `\textit{endogeneity}'
which is formally defined below. 

\begin{definition}
	Consider the LMM (\ref{EQ:model1}) and let the $j$-th covariate be denoted as $X_j$.
	\begin{itemize}
		\item We have ``unit level endogeneity" or ``\textit{level-1 endogeneity}"  when $X_j$ is correlated with the model error term $\epsilon$, 
		i.e., $Corr({X}_j, {\epsilon}_i) \neq 0$.
		We also refer to this as `\textit{error-covariate endogeniety}'.
		\item We have ``cluster level endogeneity" or ``\textit{level-2 endogeneity}" 
		when $X_j$ is correlated with some random effect $b$, i.e., $Corr({X}_j, b) \neq 0$.
		\item A variable $X_i$ is said to be `\textit{endogenous}' when it is correlated with the model error term $\epsilon$ or some random effect $b$.
		\hfill{$\square$}
	\end{itemize}
\end{definition}

The problem of endogeneity has already been extensively studied for the classical low-dimensional settings
and appropriate remedies are developed using some suitable instrumental variables (IVs);
see, among many others, Ebbes et al.~(2004, 2015), Kim and Frees (2007), Wooldridge (2010,2012), Bates et al.~(2014). 
In our ultra-high dimensional set-up, it is practically too demanding to always expect all exogeneity assumptions to hold;
in particular, the assumption regarding independence between the error and all the covarites is quite vulnerable 
and not verifiable  for extremely large $p$. 
We will see, in Section \ref{SEC:MPLE}, that the usual MPLE of the LMM parameters gets seriously affected under endogeneity; 
it also significantly increases the number of false positives in fixed effects selection.
To our knowledge, there is no literature on studying the effects of such endogeneity 
and developing appropriate remedies under high-dimensional mixed models. 
This paper aims to tackle this important problem with particular focus on level-1 endogeneity 
and to propose a new consistent selection procedure of fixed-effect variables, 
along with estimation of all parameters,  under such endogenity.

The endogeneity issue in high or ultra-high dimensional models was first considered in Fan and Liao (2014) 
under the usual regression set-up where the authors proposed a focused generalized method-of-moments (FGMM) estimator 
to consistently select and estimate the non-zero regression coefficients. 
In this paper, we will extend their FGMM approach to consistently select the important fixed effect variables under 
our ultra-high dimensional LMM set-up and then to estimate the variance parameters in a second stage. 
The proposed method is shown to satisfy the oracle 
variable selection consistency for the fixed effects even under error-covariate endogeneity. 
The overall procedure is implemented by an efficient algorithm and verified with suitable numerical illustrations.

The main contributions of the paper can be summarized as follows. 
\begin{itemize}
	\item We investigate the effect of endogeneity on the selection of fixed-effects and parameter estimation 
	in ultra-high dimensional linear mixed-effects models (LMMs). This is indeed the first such attempt for mixed models
	with exponentially increasing number of fixed-effects 
	and we prove the inconsistency of the corresponding penalized likelihood procedures under endogeneity. 
	
	\item We propose a new procedure for selecting important fixed-effects variables
	in presence of level-1 endogeneity for the ultra-high dimensional LMM. 
	Our method is based on the profiled focused generalized method of moments (PFGMM). 
	It handles the endogeneity issue through the use of appropriate IVs and 
	uses general non-concave penalties like SCAD to carry out sparse variable selection.
	The problem of unknown variance components is solved by the use of an appropriate proxy matrix.
	Our proposal is seen to produce significantly less false positives, both in simulations and in a  real data application, 
	compared to the usual penalized likelihood method of Fan and Li (2012) in the presence of endogeneity in data. 
	
	\item We rigorously prove the consistency of the estimates of fixed-effects coefficients $\boldsymbol{\beta}$
	and their oracle variable selection property under appropriate verifiable conditions. 
	Our assumptions on the penalty are completely general and cover most common non-concave penalties like either SCAD or MCP.
	The proof also allows the important selected variables to be endogenous,
	by allowing the IVs to be completely external to the regression model.
	
	\item We also prove, under appropriate conditions, an asymptotic normality result for 
	the estimates of the fixed-effects coefficients obtained by our PFGMM. 
	This will further help us to develop testing procedures in endogenous high-dimensional LMM in the future.
	
	\item  An efficient computational algorithm is also discussed along with the practical issue 
	of selecting the proxy matrix and the regularization parameter. Along with extensive numerical illustrations, 
	good-to-go suggestions for their choices are also provided which are expected to work 
	for most practical data with strong signal-to-noise ratio. 
	The (unoptimized) MATLAB code is available from the authors upon request.
	
	\item Once the important fixed-effects variables are selected consistently, 
	we also discuss and illustrate a few second stage estimation procedures to estimate 
	the variance parameters $\boldsymbol{\eta}$ along with refinements of the fixed-effects coefficients $\boldsymbol{\beta}$. 
	
	\item Although our primary focus is on level-1 endogeneity,
	finally we also briefly illustrate the effects of level-2 endogeneity 
	on our proposed PFGMM approach of variable selection. 
	Interestingly,  our proposal is seen to work consistently in most such scenarios; 
	a finding we would like to investigate theoretically in our subsequent works.
\end{itemize}

The rest of the paper is organized as follows. 
We start with the description of the usual maximum penalized likelihood 
approach and its inconsistency in the presence of endogeneity in Section \ref{SEC:MPLE}.
In Section \ref{SEC:PFGMME}, we discuss the proposed PFGMM approach with its motivation,
oracle consistency of variable selection property, asymptotic normality result
and computational aspects with numerical illustrations. Estimation of the variance parameters
in the second stage refinement are discussed and illustrated in Section \ref{SEC:Var_Est}.
The effect of level-2 endogeneity on the proposed PFGMM is examined numerically in Section \ref{SEC:level2}
and a real data application is presented in Section \ref{SEC:data}.
Finally the paper ends with brief concluding remarks in Section \ref{SEC:conclusions}.

\section{The MPLE under Endogeneity}
\label{SEC:MPLE}

Let us start with a brief description of the MPLE under the ultra-high dimensional LMM
considering the notation of Section \ref{SEC:intro}.
Using the normality of the stacked response $\boldsymbol{y}$ in our LMM (\ref{EQ:model0}),
the corresponding log-likelihood function of the parameters $(\boldsymbol{\beta}, \boldsymbol{\eta})$  turns out to be 

\begin{eqnarray}
l_n(\boldsymbol{\beta},\boldsymbol{\eta}) &=& 
-\frac{1}{2}\left[n\log(2\pi) + \log|\sigma^2 \boldsymbol{V}(\boldsymbol{\theta},\sigma^2)| 
+ \frac{1}{\sigma^2}(\boldsymbol{y}-\boldsymbol{X}\boldsymbol{\beta})^T
\boldsymbol{V}(\boldsymbol{\theta},\sigma^2)^{-1}(\boldsymbol{y}-\boldsymbol{X}\boldsymbol{\beta})\right].~~~~~~
\label{EQ:log-likelihood}
\end{eqnarray}
Adding an appropriate penalty to each component of $\boldsymbol{\beta}=(\beta_1, \ldots, \beta_p)^T$
through a penalty function $P_{n,\lambda}(\cdot)$, 
the MPLE is defined as the minimizer of the penalized objective function given by
\begin{eqnarray}
&&Q_{n,\lambda}(\boldsymbol{\beta},\boldsymbol{\eta}) = -l_n(\boldsymbol{\beta},\boldsymbol{\eta}) + \sum_{j=1}^p P_{n,\lambda}(|\beta_j|).
\label{EQ:penal_log-likelihood}
\end{eqnarray}

\noindent
With suitable regularization parameter $\lambda$, the MPLE obtained by 
the minimization of $Q_{n,\lambda}(\boldsymbol{\beta},\boldsymbol{\eta}) $ with respect to $(\boldsymbol{\beta},\boldsymbol{\eta})$ 
simultaneously selects the important (non-zero) components of $\boldsymbol{\beta}$ 
along with estimating $\boldsymbol{\eta}$ consistently. 
However, the computation is a little tricky for different penalty functions and several extensions 
have been proposed. In particular, Schelldorfer, Buhlmann and Van de Geer (2011) 
have considered the $L_1$ penalty in (\ref{EQ:penal_log-likelihood}) under high-dimensionality,
whereas Ghosh and Thoresen (2018) have extended the theory for general 
non-concave penalties under both low and high-dimensional set-ups.
An alternative two-stage approach has been proposed in Fan and Li (2012) which uses a proxy matrix 
in place of the unknown $\boldsymbol{V}(\boldsymbol{\theta}, \sigma^2)$ 
and then maximize the resulting profile likelihood of $\boldsymbol{\beta}$ only, with suitable penalizations, 
to select the important fixed effect variables;
the estimation and selection of random effect variables are considered in a second step.
Under certain assumptions, including exogeneity (no endogeneity), 
all these existing approaches to obtain the MPLE 
are shown to satisfy the oracle variable selection consistency, 
i.e., they estimate exactly the true active set (set of non-zero regression coefficients) with probability tending to one. 

We now study the effect of different types of endogeneity on the MPLE through  a numerical illustration.
Here, we use the algorithm proposed by Ghosh and Thoresen (2018) with the famous SCAD penalty 
(Antoniadis and Fan, 2001; Fan and Li, 2001); 
other existing algorithms also indicate the same behavior of the MPLE under endogeneity and are skipped for brevity. 
More illustrations are provided in later sections. 

\bigskip\noindent
\textbf{Example \ref{SEC:MPLE}.1.} 
We simulate random samples from the LMM (\ref{EQ:model1}) with $I=25$, $n_i=6$ for each $i$ (so that $n=150$), 
$p=300$, $s=5$, $q=2$ and the random effects coefficients having distribution $N(\boldsymbol{0}, \boldsymbol{\Psi}_\theta)$, 
where $\boldsymbol{\Psi}_{\theta}=Diag\{\theta_1^2, \theta_2^2\}$. 
The design matrix $\boldsymbol{X}$ has the first column as $\boldsymbol{1}$ yielding the intercept, 
and the next $(p-1)$ columns are chosen from a multivariate normal distribution with mean $\boldsymbol{0}_{p-1}$ and 
a covariance matrix having $(i,j)$-th element as $\rho^{|i-j|}$ for all $i,j=1,\ldots, p-1$;
the first two columns of $\boldsymbol{X}$ correspond to the two random effect covariates and are kept non-penalized. 
The true values of the parameters $\boldsymbol\beta$, $\sigma^2$ and $\theta_i^2$ are   
$\boldsymbol\beta=(1,2,4,3,3,0, \ldots, 0)^T$, $\sigma^2 =0.25$ and $\theta_i^2=0.56$ for $i=1,2$,
whereas $\rho=0.5$ (correlated covariates) is considered. 
The SCAD penalty is used with tuning parameter $a=3.7$
and the regularization parameter $\lambda$ is chosen by minimizing the BIC in each replication.
We replicate this process 100 times, without endogeneity, 
to compute the summary measures about the performance of the MPLE as reported in Table \ref{TAB:Sim_endo1}.

Next, to study the effects of endogenity, some covariates $X_{ij}$ are made endogenous 
with either the model error ($\epsilon_i$) or the $k$-th random effect $\boldsymbol{b}_i$, 
respectively, through the transformations 

$$
X_{ij} \leftarrow (X_{ij}+1)(\rho_{e}\epsilon_{i}+1), \mbox{ or }
X_{ij} \leftarrow (X_{ij}+1)(\rho_b b_{ik}+1), 
\mbox{ for all }i.
$$ 
These produce correlations of $\frac{\rho_{e}\sigma}{\sqrt{2\rho_{e}^2\sigma^2+1}}$ 
and $\frac{\rho_{b}\theta_k}{\sqrt{2\rho_{b}^2\theta_k^2+1}}$, respectively,
for the model error or $k$-th random-effect coefficients with the endogenous covariates. 
The summary performance measures of the resulting MPLE under such endogeneity are reported in Table \ref{TAB:Sim_endo1}
for $\rho_{e}=\rho_{b}=6$ (strong correlations of 0.688 and 0.698, respectively) and four particular sets of endogenous covariates: 
(i) Set 1: $X_6, \ldots, X_{15}$, i.e, 10 unimportant covariates are endogenous,
(ii) Set 2: $X_5, \ldots, X_{15}$, i.e, 10 unimportant covariates and one important fixed effect covariate are endogenous,
(iii) Set 3: $X_2, X_6, \ldots, X_{15}$, i.e, one important covariate that have both fixed effect component and random effect slope is endogenous along with 10 unimportant covariates,
and (iv) Set 4: $X_6, \ldots, X_{p}$, i.e, all unimportant covariates are endogenous.

\begin{table}[!th]
	\centering
	\caption{Empirical mean, SD and MSE of the parameter estimates based on penalized MLE with SCAD penalty under different types of endogeneity
		along with estimated active set size ($|S(\hat{\boldsymbol\beta})|$), number of true positives (TP) and 
		the model prediction error (PE) adjusted for the random effects 
		(the column $\boldsymbol\beta_N$ denotes the average of estimated $\beta_j$s for $j=6, \ldots,p$)}
	\resizebox{\textwidth}{!}{
		\begin{tabular}{ll|cc|c|cccccc|ccc}\hline
			Endogenous	&	covariates 	
			&	$|S(\hat{\beta})|$	&	TP	&	PE	&	$\beta_1$	&	$\beta_2$	&	$\beta_3$	&	$\beta_4$	&	$\beta_5$	&	$\beta_N$	&	$\sigma^2$	&	$\theta_1^2$	&	$\theta_2^2$	\\	\hline
			\multicolumn{14}{c}{No Endogeneity}		\\	\hline
			None	&	Mean	&	6.96	&	5.00	&	0.17	&	1.01	&	2.05	&	4.00	&	3.00	&	2.99	&	0.00	&	0.23	&	0.36	&	0.43	\\	
			&	SD	&	2.91	&	0.00	&	0.03	&	0.37	&	0.40	&	0.06	&	0.07	&	0.06	&	0.00	&	0.04	&	0.31	&	0.35	\\	
			&	MSE	&	--	&	--	&	--	&	0.1335	&	0.1604	&	0.0032	&	0.0047	&	0.0035	&	0.0000	&	0.0018	&	0.1375	&	0.1406	\\	\hline
			\multicolumn{14}{c}{Correlated with error (Level-1 endogeneity)}		\\	\hline
			Set 1	&	Mean	&	10.41	&	5.00	&	0.03	&	0.87	&	1.99	&	4.00	&	3.00	&	2.98	&	0.00	&	0.05	&	0.44	&	0.35	\\	
			&	SD	&	1.64	&	0.00	&	0.01	&	0.40	&	0.37	&	0.03	&	0.04	&	0.03	&	0.00	&	0.01	&	0.31	&	0.27	\\	
			&	MSE	&	--	&	--	&	--	&	0.1718	&	0.1329	&	0.0011	&	0.0013	&	0.0012	&	0.0000	&	0.0406	&	0.1123	&	0.1144	\\	\hline
			Set 2	&	Mean	&	10.13	&	5.00	&	0.03	&	0.90	&	2.02	&	4.00	&	2.98	&	3.03	&	0.00	&	0.05	&	0.35	&	0.39	\\	
			&	SD	&	1.95	&	0.00	&	0.01	&	0.34	&	0.38	&	0.03	&	0.03	&	0.01	&	0.00	&	0.01	&	0.28	&	0.30	\\	
			&	MSE	&	--	&	--	&	--	&	0.1222	&	0.1409	&	0.0010	&	0.0013	&	0.0008	&	0.0000	&	0.0420	&	0.1183	&	0.1187	\\	\hline
			Set 3	&	Mean	&	8.45	&	5.00	&	0.03	&	0.86	&	2.06	&	3.98	&	3.00	&	2.99	&	0.00	&	0.05	&	0.46	&	0.40	\\	
			&	SD	&	1.50	&	0.00	&	0.01	&	0.32	&	0.34	&	0.03	&	0.03	&	0.03	&	0.00	&	0.01	&	0.32	&	0.28	\\	
			&	MSE	&	--	&	--	&	--	&	0.1250	&	0.1201	&	0.0010	&	0.0009	&	0.0009	&	0.0000	&	0.0414	&	0.1138	&	0.1063	\\	\hline
			Set 4	&	Mean	&	23.45	&	5.00	&	0.01	&	0.86	&	2.00	&	3.99	&	3.00	&	2.99	&	0.00	&	0.01	&	0.49	&	0.39	\\	
			&	SD	&	4.64	&	0.00	&	0.00	&	0.38	&	0.34	&	0.01	&	0.02	&	0.02	&	0.00	&	0.00	&	0.35	&	0.28	\\	
			&	MSE	&	--	&	--	&	--	&	0.1607	&	0.1123	&	0.0003	&	0.0003	&	0.0003	&	0.0000	&	0.0586	&	0.1281	&	0.1100	\\	\hline
			\multicolumn{14}{c}{Correlated with random intercept (Level-2 endogeneity)}	\\	\hline
			Set 1	&	Mean	&	7.28	&	4.96	&	0.55	&	1.00	&	2.00	&	3.96	&	2.96	&	2.98	&	0.00	&	0.67	&	0.45	&	0.67	\\	
			&	SD	&	2.93	&	0.40	&	3.84	&	0.39	&	0.42	&	0.40	&	0.30	&	0.31	&	0.00	&	4.46	&	0.40	&	2.07	\\	
			&	MSE	&	--	&	--	&	--	&	0.1487	&	0.1781	&	0.1636	&	0.0937	&	0.0929	&	0.0000	&	19.8562	&	0.1677	&	4.2397	\\	\hline
			Set 2	&	Mean	&	7.21	&	5.00	&	0.17	&	0.99	&	2.01	&	4.00	&	3.00	&	3.00	&	0.00	&	0.23	&	0.43	&	0.44	\\	
			&	SD	&	2.54	&	0.00	&	0.03	&	0.33	&	0.32	&	0.06	&	0.06	&	0.01	&	0.00	&	0.04	&	0.37	&	0.39	\\	
			&	MSE	&	--	&	--	&	--	&	0.1075	&	0.1011	&	0.0039	&	0.0033	&	0.0001	&	0.0000	&	0.0018	&	0.1491	&	0.1672	\\	\hline
			Set 3	&	Mean	&	5.24	&	4.64	&	5.97	&	0.98	&	2.08	&	3.52	&	2.64	&	2.64	&	0.00	&	6.64	&	0.51	&	0.42	\\	
			&	SD	&	1.63	&	0.98	&	15.98	&	0.63	&	0.36	&	1.31	&	0.98	&	0.98	&	0.00	&	17.61	&	0.84	&	0.34	\\	
			&	MSE	&	--	&	--	&	--	&	0.3983	&	0.1311	&	1.9233	&	1.0834	&	1.0829	&	0.0000	&	347.9638	&	0.6949	&	0.1325	\\	\hline
			Set 4	&	Mean	&	12.47	&	4.28	&	7.66	&	0.98	&	1.62	&	3.29	&	2.46	&	2.43	&	0.00	&	8.80	&	0.43	&	4.69	\\	
			&	SD	&	6.78	&	1.54	&	16.27	&	0.36	&	0.82	&	1.55	&	1.16	&	1.15	&	0.00	&	18.63	&	0.71	&	9.74	\\	
			&	MSE	&	--	&	--	&	--	&	1.2489	&	3.2745	&	13.1795	&	6.9913	&	7.2183	&	0.1296	&	416.7764	&	0.5136	&	110.9365	\\	\hline
			\multicolumn{14}{c}{Correlated with random slope (Level-2 endogeneity)}	\\	\hline
			Set 1	&	Mean	&	7.70	&	5.00	&	0.17	&	1.02	&	1.95	&	4.00	&	3.01	&	2.98	&	0.00	&	0.24	&	0.39	&	0.42	\\	
			&	SD	&	3.58	&	0.00	&	0.03	&	0.32	&	0.38	&	0.08	&	0.07	&	0.06	&	0.00	&	0.04	&	0.31	&	0.34	\\	
			&	MSE	&	--	&	--	&	--	&	1.1992	&	3.9375	&	15.9995	&	8.6771	&	8.9100	&	0.1296	&	0.0019	&	0.1245	&	0.1353	\\	\hline
			Set 2	&	Mean	&	7.51	&	5.00	&	0.16	&	0.98	&	1.96	&	4.00	&	2.99	&	3.00	&	0.00	&	0.22	&	0.43	&	0.34	\\	
			&	SD	&	2.63	&	0.00	&	0.03	&	0.40	&	0.35	&	0.07	&	0.06	&	0.01	&	0.00	&	0.04	&	0.32	&	0.24	\\	
			&	MSE	&	--	&	--	&	--	&	1.2623	&	3.9604	&	16.0081	&	8.5702	&	8.9975	&	0.1296	&	0.0020	&	0.1208	&	0.1060	\\	\hline
			Set 3	&	Mean	&	5.28	&	4.72	&	4.75	&	0.85	&	1.94	&	3.64	&	2.73	&	2.74	&	0.00	&	5.31	&	0.51	&	0.41	\\	
			&	SD	&	1.54	&	0.90	&	14.70	&	0.63	&	0.42	&	1.15	&	0.86	&	0.87	&	0.00	&	16.29	&	0.70	&	0.34	\\	
			&	MSE	&	--	&	--	&	--	&	0.4140	&	0.1764	&	1.4431	&	0.8130	&	0.8138	&	0.0000	&	288.3149	&	0.4839	&	0.1350	\\	\hline
			Set 4	&	Mean	&	12.64	&	4.20	&	8.69	&	1.02	&	1.59	&	3.20	&	2.39	&	2.37	&	0.00	&	10.01	&	0.55	&	5.62	\\	
			&	SD	&	7.24	&	1.61	&	17.43	&	0.58	&	0.85	&	1.61	&	1.20	&	1.19	&	0.00	&	19.98	&	0.73	&	11.63	\\	
			&	MSE	&	--	&	--	&	--	&	1.4746	&	3.2622	&	12.7758	&	6.9904	&	7.0482	&	0.1296	&	490.5430	&	0.5336	&	159.4848	\\	\hline
	\end{tabular}}
	\label{TAB:Sim_endo1}
\end{table}

The major observations from Table \ref{TAB:Sim_endo1} and other similar simulations, 
not reported here for brevity, can be summarized as follows. 
\begin{itemize}
	\item Under endogeneity, we have a significant increase in false positives compared to the ideal exogenous case.
	The number of such wrongly selected fixed effect variables further	increases with 
	the strength of endogeneity and/or number of endogenous variables. 
	Such an effect is more serious for level-1 endogeneity compared to level-2 endogeneity.
	
	\item Under level-1 endogeneity, we are not expected to loose any truly significant fixed-effect variables.
	But, in some cases of level-2 endogeneity, we may loose true positives as well.
	
	\item The model prediction error is reduced in presence of level-1 endogeneity, 
	since more variables are selected in the final model. However, for level-2 endogeneity,
	the model prediction error can increase significantly when we loose the few true positives.
	
	\item The intercept is estimated with increased bias and MSE for level-1 endogeneity, 
	whereas the estimates of the other fixed-effects are affected more by level-2 endogeneity.
	
	\item The error variance also becomes severely underestimated in presence of level-1 endogeneity.
	Level-2 endogeneity has a mixed effect in this case, producing significantly
	overestimated values of $\sigma^2$ for some cases with higher degrees of endogeneity.
	
	\item As known, the random effect variances are generally underestimated even under the ideal exogenous conditions.
	The effect of endogeneity on them is not very clear but always moderate except for level-2 endogeneity
	with the full set of unimportant variables $X_6, \ldots, X_p$ (Set 4).
\end{itemize}

As our motivation is to select the important fixed-effect variables from a large pool of available candidates,
in summary, the effect of level-1 endogeneity is more serious and needs proper treatment to decrease the false positives;
on the other hand, level-2 endogeneity needs to be controlled to ensure no loss in true positives.

Having an idea of the effect of different types of endogeneity on the MPLE, 
we can now investigate this from a theoretical point of view. 
In Theorem \ref{THM:Nec_Cond}, we first present a set of necessary conditions 
for the MPLE to be consistent,  both for estimation and fixed-effects selection, in the LMM (\ref{EQ:model1}).
We will then show that at least one of these conditions do not hold under endogeneity;
hence, the MPLE is inconsistent under endogeneity.

\begin{theorem}[Necessary conditions for consistency of any sparse estimator in the LMM]\label{THM:Nec_Cond}
	Consider the LMM (\ref{EQ:model1}) where the estimation is to be performed 
	by minimizing a general loss function $L_n(\boldsymbol\beta,\boldsymbol\eta)$, need not to be the likelihood loss, 
	along with a general penalty $P_{n, \lambda}(\cdot)$. 
	Assuming sparsity of the true fixed effect coefficient $\boldsymbol{\beta}_0=(\beta_{01}, \ldots, \beta_{0p})^T$,
	let $S=\{j : \beta_{0j}\neq 0 \}$ be the (true) active set, $N=\{1, 2, \ldots, p\}\setminus S$,
	and $s=|S|$ which may or may not depend on the sample size $n$.
	Further, assume the following results hold.
	\begin{enumerate}
		\item[(C1)] $L_n(\boldsymbol\beta, \boldsymbol\eta)$ is twice differentiable with respect to its arguments 
		and the maximum  of its second derivatives at the true parameter value $(\boldsymbol\beta_0,\boldsymbol\eta_0)$ is $O_p(1)$.
		
		\item[(C2)] There is a local minimizer $(\widehat{\boldsymbol\beta}, \widehat{\boldsymbol\eta})$ 
		of the penalized objective function 
		$L_n(\boldsymbol\beta,\boldsymbol\eta)+\sum\limits_{j=1}^p P_{n,\lambda}(|\beta_j|),$ which satisfies
		$P\left(\widehat{\boldsymbol\beta}_N=\boldsymbol{0}_{p-s}\right)\rightarrow 1$, $\sqrt{s}||\widehat{\boldsymbol\beta}_S-\boldsymbol\beta_{0S}||=o_p(1)$ 
		and $||\widehat{\boldsymbol\eta}-\boldsymbol\eta_0||=o_p(1)$,
		as $n\rightarrow\infty$, where $\widehat{\boldsymbol\beta}_S$ and $\widehat{\boldsymbol\beta}_N$
		are the elements of $\widehat{\boldsymbol\beta}$ corresponding to the indices in $S$ and $N$, respectively, 
		and $\boldsymbol{\beta}_{0S}$ denotes the non-zero elements of $\boldsymbol{\beta}_0$ with indices in $S$.
		
		\item[(C3)] The penalty function $P_{n, \lambda}$ is non-negative with $P_{n, \lambda}(0)=0$, 
		$P_{n, \lambda}'(t)$ is nonincreasing on $t \in(0,u)$ for some $u>0$, and 
		$\displaystyle\lim_{n\rightarrow\infty}\lim_{t\rightarrow0+} P_{n, \lambda}'(t) = 0$.
	\end{enumerate}
	Then, for any $l\leq p$, we have
	
	\begin{eqnarray}
	\left|\frac{\partial L_n(\boldsymbol\beta_0,\boldsymbol\eta_0)}{\partial\boldsymbol\beta_l}\right|\mathop{\rightarrow}^\mathcal{P} 0.
	\label{EQ:Nec_cond}\vspace{-2cm}
	\end{eqnarray}
\end{theorem}

The proof of this theorem has been given in the Online Supplementary material.
Here, it is important to note that Theorem \ref{THM:Nec_Cond} is established without any reference 
to the exogeneity or endogeneity conditions. It presents a necessary condition (\ref{EQ:Nec_cond}) 
for the parameter estimates to be consistent, as in (C2), for a general class of loss functions satisfying (C1) 
and the  penalties satisfying (C3). Since it is a necessity result, the rate of consistency in (C2) is not important.
Further, Condition 3 about the penalty function is indeed the same as the one used in Theorem 2.1 of Fan and Liao (2014) 
and is quite general; it is satisfied by most common penalties including $L_1$, SCAD or MCP by appropriately choosing 
the sequence of regularization parameter $\lambda=\lambda_n$.
Thus, as in Fan and Liao (2014), our result in Theorem \ref{THM:Nec_Cond} rather provides 
a necessary condition (\ref{EQ:Nec_cond}) on the loss function for a large class of useful penalty functions.
Since (C1) always holds for the likelihood loss, 
if (\ref{EQ:Nec_cond}) is not satisfied then all the consistency results in (C2) cannot hold for the resulting MPLE.
It is known that (C2) and hence (\ref{EQ:Nec_cond}) must hold for the MPLE under exogeneity.
In the following theorem, we will show that (\ref{EQ:Nec_cond}) fails to hold under any sort of endogeneity indicating 
the inconsistency of the MPLE, in at least one aspect; the proof is given in the Online Supplementary material.

\begin{theorem}[Inconsistency of the MPLE in Endogenous LMM]
	Consider the LMM (\ref{EQ:model1})	with the likelihood loss given by (\ref{EQ:log-likelihood}), in negative,
	and the $P_{n,\lambda}(t)$ satisfying Condition (C3) of Theorem \ref{THM:Nec_Cond}.
	Suppose that at least one $X$ in at least one group $i$ is endogenous (level-1 or level-2)
	and that the $X$ and the model error $\epsilon$ both have finite 4th order moments.
	If $(\widehat{\boldsymbol\beta}, \widehat{\boldsymbol\eta})$  denotes a (local) MPLE such that  
	$||\widehat{\boldsymbol{\eta}}-\boldsymbol{\eta}_0||=o_p(1)$, 
	then either 
	$$
	\limsup\limits_{n\rightarrow\infty} P(\widehat{\boldsymbol{\beta}}_N=0) < 1, 
	~~\mbox{ or }~ ||\widehat{\boldsymbol{\beta}}_S-\boldsymbol{\beta}_{0S}||\neq o_p(1),
	$$
	where $\widehat{\boldsymbol{\beta}}_S$, $\widehat{\boldsymbol{\beta}}_N$, $\boldsymbol{\beta}_{0S}$
	and $\boldsymbol{\eta}_0$ are as defined in Theorem \ref{THM:Nec_Cond} for the MPLE.
\end{theorem}


%
\section{Focused Selection of Fixed Effect Variables under Level-1 Endogeneity}\label{SEC:PFGMME}

Consider the LMM set-up as described in Section \ref{SEC:intro}. 
We now propose a new extension of the MPLE of Fan and Li (2012),
that will lead to consistent oracle selection of important fixed effect variables, 
using the FGMM approach with non-concave penalization.
The FGMM loss function, as initially proposed  by Fan and Liao (2014) in the context of high-dimensional linear regression, 
simultaneously performs sparse selection and applies the IV method against endogeneity.
The IV method basically assumes the availability of a vector of observable \textit{instrumental variables} $\boldsymbol{W}$
which is correlated with the covariates $\boldsymbol{X}$ but uncorrelated with the model error, i.e.,
$E[\epsilon|\boldsymbol{W}]=0.$
The choice of a proper IV (s) helps us to tackle different statistical problems;
they are often chosen as a function of the covariates or even a subset of $\boldsymbol{X}$
and hence the above condition can be easily verified through some simple moment conditions.
As noted earlier, the IV technique is seen to be extremely useful to address the endogeneity issues 
in classical low-dimensional LMM;
see  Hall and Horowitz (2005), Wooldridge (2010), Lin et al.~(2015), Chesher and Rosen (2017) 
for some recent IV methods.

\subsection{The Profiled Focused GMM (PFGMM) with non-concave penalization}

Under the LMM set-up considered in this paper, keeping consistent with the many real-life applications, 
we have assumed that the number of random effects are small enough so that their individual analysis is possible in the classical sense.
Hence, we assume that the matrix $\boldsymbol{\Psi}_{\boldsymbol{\theta}}$ is positive definite (pd).
Let us first assume, for the time being, that the variance parameter $\boldsymbol{\eta}= (\boldsymbol{\theta}^T, \sigma^2)^T$ is known. 
Then, based on (\ref{EQ:log-likelihood}), the likelihood of the only parameter $\boldsymbol{\beta}$ becomes
\begin{equation}
L_{\mbox{profile}}(\boldsymbol{\beta}) \propto \exp \left\{- \frac{1}{2\sigma^2}(\boldsymbol{y}-\boldsymbol{X}\boldsymbol{\beta})^T
\boldsymbol{V}(\boldsymbol{\theta},\sigma^2)^{-1}(\boldsymbol{y}-\boldsymbol{X}\boldsymbol{\beta})\right\}.
\label{EQ:prof-liklihood}
\end{equation}
Note that, this is also the profiled likelihood of $\boldsymbol{\beta}$ 
obtained by  substituting the MLE of the random effect vector $\boldsymbol{b}=(\boldsymbol{b}_1^T, \ldots, \boldsymbol{b}_I^T)^T$, given $\boldsymbol{\beta}$, 
in the joint likelihood of $\boldsymbol{y}=(\boldsymbol{y}_1^T, \ldots, \boldsymbol{y}_I^T)^T$ and $\boldsymbol{b}$ given covariates;
see Fan and Li (2012) for details. A penalized version of this profile likelihood (in logarithm) can be maximized 
for sparse selection of the fixed effects and estimation of the corresponding coefficients; 
Fan and Li (2012) have suggested to use a suitable proxy matrix for the unknown $\boldsymbol{V}(\boldsymbol{\theta},\sigma^2)$.

Now, in the presence of endogeneity, we need to additionally apply the IV method to achieve consistency.
Let us again assume, for the time being, that  $\boldsymbol{\eta}= (\boldsymbol{\theta}^T, \sigma^2)^T$ 
and hence  $\boldsymbol{V}(\boldsymbol{\theta},\sigma^2)$ is known. Let us define the transformed variables

$$
\boldsymbol{y}^\ast = \boldsymbol{V}(\boldsymbol{\theta},\sigma^2)^{-1/2}\boldsymbol{y}, ~~~~~~
\boldsymbol{X}^\ast = \boldsymbol{V}(\boldsymbol{\theta},\sigma^2)^{-1/2}\boldsymbol{X}, ~~~~~~
\boldsymbol{\epsilon}^\ast = \boldsymbol{V}(\boldsymbol{\theta},\sigma^2)^{-1/2}\boldsymbol{\epsilon}.
$$
Then we have $\boldsymbol{\epsilon}^\ast \sim N_n(\boldsymbol{0}, \sigma^2\boldsymbol{I}_p)$ 
and hence $\boldsymbol{y}^\ast \sim N_n(\boldsymbol{X}^\ast\boldsymbol{\beta}, \sigma^2\boldsymbol{I}_p)$.
Therefore, the profile likelihood of $\boldsymbol{\beta}$, given in (\ref{EQ:prof-liklihood}), 
under the LMM (\ref{EQ:model1}) is also the ordinary likelihood 
of $\boldsymbol{\beta}$ under the following linear regression model in the transformed space:

\begin{eqnarray}
\boldsymbol{y}^\ast = \boldsymbol{X}^\ast\boldsymbol{\beta} + \boldsymbol{\epsilon}^\ast, ~~~~ \boldsymbol{\epsilon}^\ast \sim N_n(\boldsymbol{0}, \sigma^2\boldsymbol{I}_p).
\label{EQ:model_transf}
\end{eqnarray}

\noindent
Under level-1 endogeneity in the LMM (\ref{EQ:model1}), 
we also have endogeneity in the transformed regression (\ref{EQ:model_transf})
with  $E \left[\boldsymbol{y}^\ast - \boldsymbol{X}^\ast\boldsymbol{\beta}| \boldsymbol{X}^\ast \right]\neq \boldsymbol{0}$.
Noting that (\ref{EQ:model_transf}) is exactly the same model as considered by Fan and Liao (2014), 
our idea is to apply their FGMM approach to this transformed model in the transformed space 
and then go back to the original space of the data to achieve our goal of 
fixed effects selection in the LMM (\ref{EQ:model1}) with endogeneity. 
This would have been straightforward if the original data were independent and identically distributed (iid) and  
$\boldsymbol{V}(\boldsymbol{\theta},\sigma^2)$ was known, but none of these conditions hold in practice. 
Hence, we need appropriate \textit{non-trivial extensions} to handle 
the implementation and theoretical derivations. 
Let us start with defining our proposed loss function.

Note that the components of $\boldsymbol{\epsilon}^\ast$ are iid and 
those of $\boldsymbol{y}^\ast$ are independent with the same variance but different means. 
Let us denote the corresponding random variables in the transformed space by $\epsilon^\ast$ and $Y^\ast$, respectively. 
Then, obtaining a consistent solution under endogeneity is based on 
the availability of an appropriate set of observable instrumental variables $\boldsymbol{W}^\ast$ 
in the transformed space such that 
$$
E\left[\epsilon^\ast | \boldsymbol{W}^\ast  \right] =0.
$$
Fan and Liao (2014) achieved variable selection consistency under endogeneity 
through over-identification via the use of two sets of sieve functions (Chen, 2007), say, 
$\boldsymbol{F}^\ast = (f_1(\boldsymbol{W}^\ast), \ldots, f_p(\boldsymbol{W}^\ast))^T$
and  $\boldsymbol{H}^\ast = (h_1(\boldsymbol{W}^\ast), \ldots, h_p(\boldsymbol{W}^\ast))^T$
where $f_j$ and $h_j$ are scalar functions. Letting $S$ denote the index set of true non-zero coefficients,
the above condition of IV implies that, for $\boldsymbol{\beta}_S = \boldsymbol{\beta}_{0S}$,
we have the following set of over-identified equations:

\begin{equation}
E\left[(Y^\ast - \boldsymbol{X}_S^\ast\boldsymbol{\beta}_S)\boldsymbol{F}_S^\ast\right] = \boldsymbol{0},
~~~~
E\left[(Y^\ast - \boldsymbol{X}_S^\ast\boldsymbol{\beta}_S)\boldsymbol{H}_S^\ast\right] = \boldsymbol{0}.
\label{EQ:IV_eq_trans}
\end{equation}

\noindent
Under these conditions, Fan and Liao (2014) have proposed to consider the FGMM loss function 
\vspace{-1.5cm}
\begin{eqnarray}
L_n(\boldsymbol{\beta}) &=& 
\left[\frac{1}{n}\sum_{i=1}^n (Y_i^\ast - \boldsymbol{X}_i^\ast\boldsymbol{\beta})\boldsymbol{\Pi}_i^\ast(\boldsymbol{\beta})\right]^T
\boldsymbol{J}(\boldsymbol{\beta})
\left[\frac{1}{n}\sum_{i=1}^n (Y_i^\ast - \boldsymbol{X}_i^\ast\boldsymbol{\beta})\boldsymbol{\Pi}_i^\ast(\boldsymbol{\beta})\right]\\
&=& \left[\frac{1}{n}\boldsymbol{\Pi}^\ast(\boldsymbol{\beta}) (\boldsymbol{y}^\ast - \boldsymbol{X}^\ast\boldsymbol{\beta})\right]^T
\boldsymbol{J}(\boldsymbol{\beta})
\left[\frac{1}{n}\boldsymbol{\Pi}^\ast(\boldsymbol{\beta}) (\boldsymbol{y}^\ast - \boldsymbol{X}^\ast\boldsymbol{\beta})\right],
\label{EQ:FGMM_loss_transf}
\end{eqnarray}

\noindent
where $\boldsymbol{\Pi}_i^\ast(\boldsymbol{\beta}) =(\boldsymbol{F}_i^\ast(\boldsymbol{\beta})^T, \boldsymbol{H}_i^\ast(\boldsymbol{\beta})^T)^T$ for all $i$ 
and $\boldsymbol{J}(\boldsymbol{\beta})$ is a diagonal weight matrix 
with non-zero weights corresponding only to the non-zero components of $\boldsymbol{\beta}$.
In particular, the non-zero weights of $j$ components can be chosen as the inverse of the estimated variances of 
$f_j(\boldsymbol{W}^\ast)$ and $h_j(\boldsymbol{W}^\ast)$, respectively.
Then, a consistent solution of the transformed problem can be obtained by minimizing the penalized FGMM loss function;
see Fan and Liao (2014) for details.

Now, let us look back at our original problem under the LMM (\ref{EQ:model1})
and map the FGMM loss function (\ref{EQ:FGMM_loss_transf}) back into our data space to get a clear idea for this case.
Note that, through an inverse transformation, 
we can assume $\boldsymbol{\Pi}^\ast(\boldsymbol{\beta}) = \boldsymbol{\Pi}(\boldsymbol{\beta})\boldsymbol{V}(\boldsymbol{\theta},\sigma^2)^{-1/2}$
for some IV $\boldsymbol{\Pi}$ in the data space. 
Hence the FGMM loss function for our mixed model set-up turns out to have the form 
\begin{eqnarray}
L_n(\boldsymbol{\beta}) &=& 
\left[\frac{1}{n}\boldsymbol{\Pi}(\boldsymbol{\beta})\boldsymbol{V}(\boldsymbol{\theta},\sigma^2)^{-1} 
(\boldsymbol{y} - \boldsymbol{X}\boldsymbol{\beta})\right]^T
\boldsymbol{J}(\boldsymbol{\beta}) \left[\frac{1}{n}\boldsymbol{\Pi}(\boldsymbol{\beta})\boldsymbol{V}(\boldsymbol{\theta},\sigma^2)^{-1} 
(\boldsymbol{y} - \boldsymbol{X}\boldsymbol{\beta})\right].
\label{EQ:FGMM_loss}
\end{eqnarray}

\noindent
Note that, in practice with mixed models, we can not directly minimize this FGMM loss function or its penalized version
because it depends on the unknown variance parameters through $\boldsymbol{V}(\boldsymbol{\theta},\sigma^2)$.
To avoid this problem, we follow the approach of Fan and Li (2012) and propose 
to use $\widetilde{\boldsymbol{V}}_z = \left[\boldsymbol{I}_n + \boldsymbol{Z}^T\mathcal{M}\boldsymbol{Z}\right]$ 
in place of $\boldsymbol{V}(\boldsymbol{\theta},\sigma^2)$, 
where $\mathcal{M}$ is some suitable proxy matrix for the unknown variance component matrix $\sigma^{-2}\boldsymbol{\Psi}_{\boldsymbol{\theta}}$.
Therefore, we finally minimize, with respect to $\boldsymbol{\beta}$, the penalized objective function 
\begin{eqnarray}
Q_n(\boldsymbol{\beta}) &=& L_n^P(\boldsymbol{\beta}) + \sum_{j=1}^p P_{n,\lambda}(|\beta_j|),
\label{EQ:PFGMM_loss_penal}~\\
\mbox{where }~~
L_n^P(\boldsymbol{\beta}) &=& 
\left[\frac{1}{n}\boldsymbol{\Pi}(\boldsymbol{\beta})\widetilde{\boldsymbol{V}}_z^{-1} (\boldsymbol{y} - \boldsymbol{X}\boldsymbol{\beta})\right]^T
\boldsymbol{J}(\boldsymbol{\beta}) \left[\frac{1}{n}\boldsymbol{\Pi}(\boldsymbol{\beta})\widetilde{\boldsymbol{V}}_z^{-1} 
(\boldsymbol{y} - \boldsymbol{X}\boldsymbol{\beta})\right].
\label{EQ:PFGMM_loss}
\end{eqnarray}

\noindent
We will refer to $L_n^P$ as the profiled Focused GMM (PFGMM) loss function based on its link to the profile likelihood.
If we would have used $\boldsymbol{V}(\boldsymbol{\theta},\sigma^2)$ for known variance parameters, 
the asymptotic consistency results for the resulting estimator would have followed directly from the results of Fan and Liao (2014).
However, we will here prove that, even using the proxy matrix $\widetilde{\boldsymbol{V}}_z$,
we can still achieve variable selection consistency under the linear mixed  model
provided the proxy matrix is not very far away from the truth; 
we will present the rigorous proof along with necessary assumptions in the next subsection.

\subsection{Oracle Variable Selection Consistency }

Consider the set-up of the previous subsection and assume that the true parameter value 
$\boldsymbol{\beta}_0 = (\boldsymbol{\beta}_{0S}^T, \boldsymbol{0})^T$ is the unique solution of 
the set of over-identified IV equations in (\ref{EQ:IV_eq_trans}), 
where the non-zero component vector $\boldsymbol{\beta}_{0S}\in\mathbb{R}^s$. 
Further, we need the following sets of assumptions.

\bigskip
\noindent
\textbf{Assumptions on the penalty (P):}\\
The general penalty function $P_{n,\lambda}(t): [0, \infty) \rightarrow \mathbb{R}$ satisfies
\begin{itemize}
	\item[(P1)] $P_{n,\lambda}(t)$ is concave and non-decreasing on $[0,\infty)$, with $P_{n,\lambda}(0) = 0$,
	
	\item[(P2)] $P_{n,\lambda}(t)$ has continuous derivative $P_{n,\lambda}'(t)$ on $(0,\infty)$, with 
	$\sqrt{s} P_{n,\lambda}'(d_n) = o(d_n)$, \\where 
	$d_n=\frac{1}{2}\min\{|\beta_{0j}|:\beta_{0j}\neq 0, ~j=1, \ldots, p\}$  denotes the strength of the signal,
	
	
	\item[(P3)] There exists a constant $c>0$ such that 
	$\sup_{\boldsymbol\beta\in B(\boldsymbol\beta_{S_0}, cd_n)} \zeta(\boldsymbol\beta) = o(1)$, where
	\begin{equation}
	\zeta(\boldsymbol\beta) = \limsup_{\epsilon \rightarrow 0+} \max_{j\leq s} \sup_{t_1<t_2: (t_1,t_2)\in (|\beta_j|-\epsilon, |\beta_j|+\epsilon)}
	- \left[\frac{P_{n,\lambda}(t_2) - P_{n,\lambda}(t_1)}{t_2-t_1}\right].
	\label{EQ:zeta_beta}
	\end{equation}
\end{itemize}

\vspace{-0.5cm}
It is worthwhile to note that Conditions (P1)--(P3) are quite standard in high-dimensional analysis
and used by several authors including Fan and Liao (2014).  These are satisfied by a large class of folded-concave penalties
including $L_q$ with $q\leq 1$, hard-thresholding, SCAD and MCP for appropriately chosen tuning parameters. 
Also $\zeta(\boldsymbol\beta) \geq 0$ for any $\boldsymbol{\beta}\in \mathbb{R}^s$, by the concavity of the penalty functions. 
Condition (P2) is related to the signal strength, 
on which we need the following additional assumptions depending on the dimension of the problem;
these are needed to ensure variable selection consistency 
and are satisfied also by properly chosen SCAD and MCP penalties for strong signal $d_n$ and small $s\ll n$.

\noindent
\textbf{Assumptions on the dimension and signal strength  (A):}
\begin{itemize}
	\item[(A1)] $P_{n,\lambda}'(d_n) = o(1/\sqrt{ns})$, $sP_{n,\lambda}'(d_n) +s\sqrt{\log p/n} + s^3 \log s /n = o(P_{n,\lambda}'(0^{+}))$,
	$P_{n,\lambda}'(d_n)s^2 = O(1)$.	
	
	\item[(A2)]  $s\sqrt{\log p /n}=o(d_n)$ and 
	$\displaystyle\sup_{||\boldsymbol\beta-\boldsymbol\beta_{S0}||\leq d_n/4} \zeta(\boldsymbol\beta) = o(1/\sqrt{s\log p}).$
\end{itemize}

Next we assume the following conditions on the instrumental variables $\boldsymbol{F}^\ast$
and $\boldsymbol{H}^\ast$ with the notation $F_j^\ast = f_j(\boldsymbol{W}^\ast)$ and 
$H_j^\ast = h_j(\boldsymbol{W}^\ast)$ for $j=1, 2, \ldots, p$.
These are motivated from Fan and Liao (2014) and 
similar justifications hold for their selection; see Remark 4.1 there.
We use the notations $\lambda_{\min}$ and $\lambda_{\max}$ to  denote the smallest 
and the largest eigenvalues, respectively.

\noindent
\textbf{Assumptions on the Instruments (I):}
\begin{itemize}
	\item[(I1)] There exists $b_1, b_2, r_1, r_2>0$ such that 
	$$
	\max\limits_{l\leq p} P(|F_l^\ast|>t) \leq e^{-(\frac{t}{b_1})^{r_1}},
	~~~~
	\max\limits_{l\leq p} P(|H_l^\ast|>t) \leq e^{-(\frac{t}{b_2})^{r_2}},
	~~\mbox{for any } t>0.
	$$
	
	\item[(I2)] Var$\left({F}_j^\ast\right)$ and Var$\left({H}_j^\ast\right)$  are bounded away from 
	both zero and infinity	uniformly in $j=1, \ldots, p$ and $p \geq 1$.

	\item[(I3)] $\min\limits_{j\in S} \mbox{Var}\left((Y^\ast - \boldsymbol{X}^\ast\boldsymbol{\beta}_0){F}_j^\ast\right)$ 
	and $\min\limits_{j\in S} \mbox{Var}\left((Y^\ast - \boldsymbol{X}^\ast\boldsymbol{\beta}_0){H}_j^\ast\right)$
	are bounded away from zero.
	
	\item[(I4)] There exist constants $C_1, C_2>0$ such that $\lambda_{\max}(\boldsymbol{A}\boldsymbol{A}^T)<C_1$ 
	and $\lambda_{\min}(\boldsymbol{A}\boldsymbol{A}^T)>C_2$, where
		
	$$
	\boldsymbol{A} = \lim\limits_{n\rightarrow\infty}\frac{1}{n}\boldsymbol{\Pi}(\boldsymbol{\beta}_{0S})\widetilde{\boldsymbol{V}}_z^{-1}\boldsymbol{X}_S.
	$$
	%
	\item[(I5)] There exists a constant $C>0$ such that $\lambda_{\min}(\boldsymbol{\Upsilon})>C$, where
	$$\boldsymbol{\Upsilon} = \lim\limits_{n\rightarrow\infty}\frac{\sigma^2}{n}\boldsymbol{\Pi}(\boldsymbol{\beta}_0)\widetilde{\boldsymbol{V}}_z^{-1}
	\boldsymbol{V}(\boldsymbol{\theta},\sigma^2)\widetilde{\boldsymbol{V}}_z^{-1}\boldsymbol{\Pi}(\boldsymbol{\beta}_0)^T.	
	$$
\end{itemize}

Note that, Assumptions (I4) and (I5) additionally depend on the choice of proxy matrix $\mathcal{M}$,
which appear in $\widetilde{\boldsymbol{V}}_z$. 
This $\widetilde{\boldsymbol{V}}_z$ is the key to handle the presence of random effects in the loss function
by substituting the unknown $V(\boldsymbol{\theta}, \sigma^2)$. 
However, we do not need $\widetilde{\boldsymbol{V}}_z$ to be consistent for  $V(\boldsymbol{\theta}, \sigma^2)$
in our derivations; it is sufficient that the proxy matrix $\mathcal{M}$ is close to 
$\sigma^{-2}\boldsymbol{\Psi}_{\boldsymbol{\theta}}$ in the sense of the following assumption.

\noindent
\textbf{Assumptions on the Proxy Matrix  (M):}
\begin{itemize}
	\item[(M1)] $\lambda_{\min}\left[C_1\mathcal{M} - \sigma^{-2}\boldsymbol{\Psi}_{\boldsymbol{\theta}} \right] \geq 0$
	and
	$\lambda_{\min}\left[C_1\log n(\sigma^{-2}\boldsymbol{\Psi}_{\boldsymbol{\theta}}) - \mathcal{M}\right] \geq 0$,
	for some $C_1>1$.
	\item[(M2)] $\displaystyle\max_{j\notin S}||\boldsymbol{A}_j||\sqrt{\log s/n} = o(P_{n,\lambda}(0^{+}))$,
	where $\boldsymbol{A}_j$ denotes the $j$-th column of the matrix $\boldsymbol{A}$ defined in Assumption (I4).
\end{itemize}

Then we have the following main theorem. For simplicity in presentation, 
we have deferred its proof to the Online Supplementary material.

\begin{theorem}
	Consider the set-up of LMM (\ref{EQ:model1}) with the true parameter value being $(\boldsymbol{\beta}_0, \boldsymbol{\eta}_0)$
	Assuming sparsity of $\boldsymbol{\beta}_0=(\beta_{01}, \ldots, \beta_{0p})^T$,
	let $S=\{j : \beta_{0j}\neq 0 \}$ be the (true) active set with size $s=|S|$ and $N=\{1, 2, \ldots, p\}\setminus S$.
	Suppose $s^3\log p = o(n)$ and Assumptions (P), (A), (I) and (M) hold.
	Then, there exists a  local minimizer 
	$\widehat{\boldsymbol{\beta}} =(\widehat{\beta}_1, \ldots, \widehat{\beta}_p)^T$
	of the PFGMM objective function $Q(\boldsymbol\beta)$ in (\ref{EQ:PFGMM_loss_penal}) that satisfies the following properties.
	\begin{enumerate} 
		\item[a)] $\displaystyle\lim_{n\rightarrow\infty} P(\widehat{\boldsymbol{\beta}}_N =\boldsymbol{0})=1$,
		where $\widehat{\boldsymbol\beta}_N$ corresponds to the elements of $\widehat{\boldsymbol\beta}$ with indices in $N$.
		
		\item[b)] If $\widehat{S} = \{j\leq p : \widehat{{\beta}}_j \neq 0\}$ denotes the estimated active set, 
		then $\displaystyle\lim_{n\rightarrow\infty} P(\widehat{S}=S)=1$.
		
		\item[c)] For any unit vector $\boldsymbol\alpha\in\mathbb{R}^s$, 
		$
		\sqrt{n}\boldsymbol\alpha^t \boldsymbol{\Gamma}^{-1/2}\boldsymbol{\Sigma} (\widehat{\boldsymbol{\beta}}_S-\boldsymbol\beta_{0S}) 
		\displaystyle\mathop\rightarrow^\mathcal{D} N(0,1),
		$
		where $\widehat{\boldsymbol\beta}_S$ corresponds to the elements of $\widehat{\boldsymbol\beta}$ with indices in $S$, $\boldsymbol{\beta}_{0S}$ denotes the non-zero elements of $\boldsymbol{\beta}_0$ with indices in $S$, 
		$\boldsymbol{\Gamma} = 4
		\boldsymbol{A}\boldsymbol{J}(\boldsymbol{\beta}_0)\boldsymbol{\Upsilon}\boldsymbol{J}(\boldsymbol{\beta}_0)\boldsymbol{A}^T$
		and $\boldsymbol{\Sigma} = 2 \boldsymbol{A}\boldsymbol{J}(\boldsymbol{\beta}_0)\boldsymbol{A}^T$.
		\item[d)] 	In addition, the local minimizer $\widehat{\boldsymbol{\beta}}$ 
		is strict with probability arbitrarily close to one for all sufficiently large $n$.
	\end{enumerate}
	\label{THM:main}
\end{theorem}

\begin{remark}
	Although we have three types of unknown parameters $(\boldsymbol{\beta}, \boldsymbol{\theta}, \sigma^2)$ 
	in the LMM (\ref{EQ:model1}), 
	the PFGMM loss function depends only on the fixed-effects coefficient parameter $\boldsymbol{\beta}$.
	Thus, the minimization of the PFGMM objective function $Q(\boldsymbol\beta)$ in (\ref{EQ:PFGMM_loss_penal}) 
	only produce an estimate $\widehat{\boldsymbol{\beta}}$ of $\boldsymbol{\beta}$ which, in turn,
	also selects the important fixed-effects associated with the non-zero coefficients in $\widehat{\boldsymbol{\beta}}$ 
	due to the use of a sparse non-concave penalty function. 
	Once $\widehat{\boldsymbol{\beta}}$ is obtained, the estimation of the other variance parameters 
	$\boldsymbol{\eta} =(\boldsymbol{\theta}, \sigma^2)$ needs to be done in a second stage,
	which is described later in Section \ref{SEC:Var_Est}.
\end{remark}

\begin{remark}\label{REM:level2}
	It is worthwhile to note that, although we have proposed the PFGMM approach considering level-1 endogeneity 
	in the LMM (\ref{EQ:model1}), its oracle consistency results in Theorem \ref{THM:main} is not hampered 
	by the presence of level-2 endogeneity. This is because the transformed model (\ref{EQ:model_transf})
	and hence the PFGMM loss  function in (\ref{EQ:PFGMM_loss}) do not involve the random effects $\boldsymbol{b}$
	if the proxy matrix is chosen appropriately. However, the required assumptions might become stricter 
	if endogeneity is present also in the associated random effect covariates $\boldsymbol{Z}$. 
	Therefore, we expect the proposed PFGMM approach to work well in selecting important fixed-effect variables
	even under level-2 endogeneity in well-specified LMM; 
	we will further illustrate this aspect empirically through simulations in Section \ref{SEC:level2}.
\end{remark}


\subsection{Computational Aspects}
\label{SEC:PFGMM_Computation}

For implementing the proposed PFGMM algorithm, 
one can follow the same algorithm as used by Fan and Liao (2014)
on the transformed variables $\boldsymbol{y}^\ast$ and $\boldsymbol{X}^\ast$.
But, these transformed variables leading to the loss function in (\ref{EQ:FGMM_loss})
are not known, so  we need to use the proxy matrix and the approximated loss given in (\ref{EQ:PFGMM_loss}).
So, given a proxy matrix $\mathcal{M}$, we first compute the matrix $\widetilde{\boldsymbol{V}}_z^{-1}$
and its square root $\widetilde{\boldsymbol{V}}_z^{-1/2}$.
For computation of the matrix square root we use the Blocked Schur algorithm 
as developed by Deadman et al.~(2013); its implementation can be found in standard statistical softwares like
MATLAB or R (function named `\textit{sqrtm}' in both).
Then, the approximation of the transformed variables are computed as 
$\widetilde{\boldsymbol{y}}^\ast = \widetilde{\boldsymbol{V}}_z^{-1/2}\boldsymbol{y}$
and $\widetilde{\boldsymbol{X}}^\ast = \widetilde{\boldsymbol{V}}_z^{-1/2}\boldsymbol{X}$.
Following this, the FGMM loss function based on $\widetilde{\boldsymbol{y}}^\ast$ 
and $\widetilde{\boldsymbol{X}}^\ast$ is nothing but the proposed loss in (\ref{EQ:PFGMM_loss})
and hence we can next device an algorithm following the Fan and Liao (2014) approach.
The minimization of the resulting penalized PFGMM objective function
is done through the iterative coordinate algorithm applied to a smoothed version of 
the non-smooth PFGMM minimization problem. 
More details and justifications can be found in Fan and Liao (2014);
for brevity, we only present the crucial considerations regarding
the choice of proxy matrix and the choice of regularization parameter $\lambda$ in our context. 

\bigskip
\noindent
\textbf{On the Choice of Proxy matrix:}\\
The choice of proxy matrix $\mathcal{M}$ is not straightforward from the assumed conditions
but some light can be shed following the discussions in Fan and Li (2012, Section 2.3).
In particular, assuming standard non-singularity conditions involving the random effects covariates $\boldsymbol{Z}$, 
one possible choice of $\mathcal{M}$ which can be obtained for large $n$ is $\log(n)$ times the identity matrix.
We have used this particular proxy matrix for all our empirical illustrations in the present paper.

\bigskip
\noindent
\textbf{On the Choice of $\lambda$:}\\
Although for the regression modeling considered in Fan and Liao (2014), 
the regularization parameter $\lambda$ can be chosen by cross-validation 
and hence can also be used in connection with any loss function other than likelihood-loss (like the FGMM loss),
it is not ideal to apply the cross-validation technique in case of mixed models.  
In likelihood based estimation and variable selection in high or ultra-high dimensional LMMs, 
the usual proposal is to choose $\lambda$ corresponding to the minimum value of the BIC given by 
(Schelldorfer et al., 2011; Delattre et al., 2014; Ghosh and Thoresen, 2018)
$$
\mbox{BIC}(\lambda) = -2l_n\left(\widehat{\boldsymbol{\beta}}, \widehat{\boldsymbol{\eta}}\right) + [|\widehat{S}|+dim(\lambda)]\log n.
$$
When we are using the proposed PFGMM loss function to estimate $\boldsymbol{\beta}$ 
in the ultra-high dimensional mixed model, one can define a natural extension of BIC as 
$$
\mbox{ExBIC}(\lambda) = -2L_n^P\left(\widehat{\boldsymbol{\beta}}_\lambda^P\right) + |\widehat{S}|\log n,
$$
where $\widehat{\boldsymbol{\beta}}_\lambda^P$ is the estimate of $\boldsymbol{\beta}$ 
obtained through the proposed PFGMM approach with regularization parameter $\lambda$.
But, in this context, it may be questionable if the above formulations provide the 
correct penalty and this clearly needs further detailed investigation.   
However, it has been observed that the simple choice of $\lambda=0.1$,
as suggested in Fan and Liao (2014), works sufficiently well for all our numerical studies.

\begin{figure}[!b]
	\centering
	\subfloat[Endogenous variables: Set 1]{
		\includegraphics[width=0.45\textwidth]{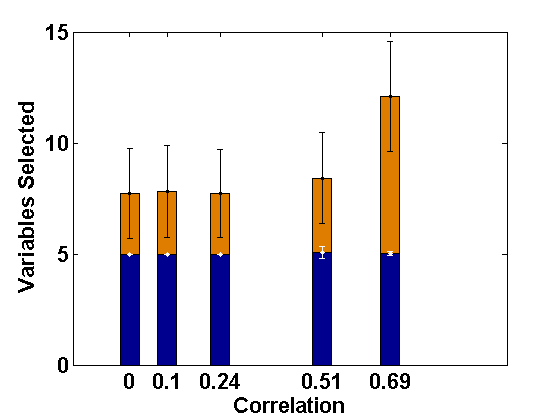}
		\label{FIG:9size_unknown_30_0}}
	~ 
	\subfloat[Endogenous variables: Set 2]{
		\includegraphics[width=0.45\textwidth]{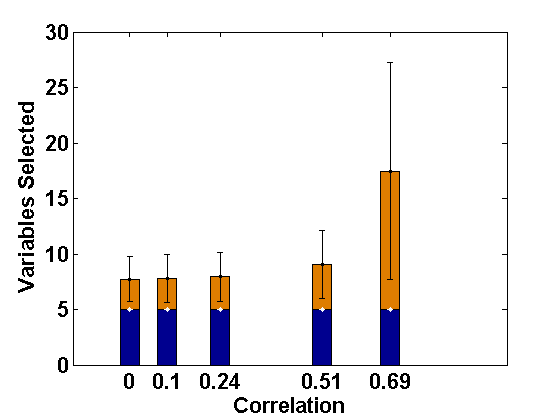}
		\label{FIG:9size_unknown_30_05}}
	\\
	\subfloat[Endogenous variables: Set 3 ]{
		\includegraphics[width=0.45\textwidth]{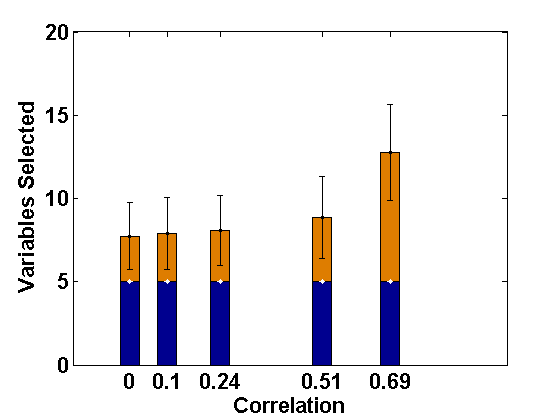}
		\label{FIG:9size_unknown_50_0}}
	~ 
	\subfloat[Endogenous variables: Set 4]{
		\includegraphics[width=0.45\textwidth]{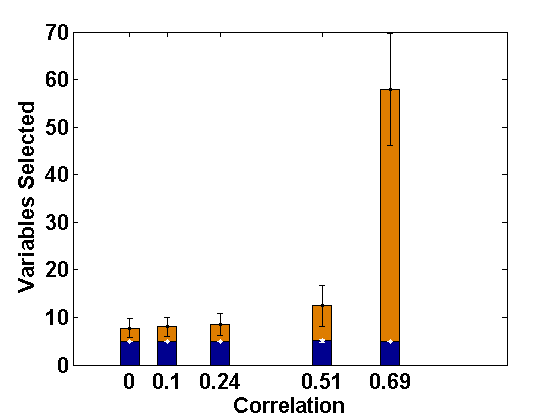}
		\label{FIG:9size_unknown_50_05}}
	\caption{Sizes of the estimated active set by the PFGMM (blue) method and the PLS method (blue+orange)
		for different extents of endogeneity and  correlated covariates 
		(the standard errors are shown by respective error bars)}
	\label{FIG:ActiveSet_X05}
\end{figure}

\subsection{Empirical illustrations}
\label{SEC:EX_PFGMM}

\begin{figure}[!h]
	\centering
	\subfloat[Endogenous variables: Set 1]{
		\includegraphics[width=0.45\textwidth]{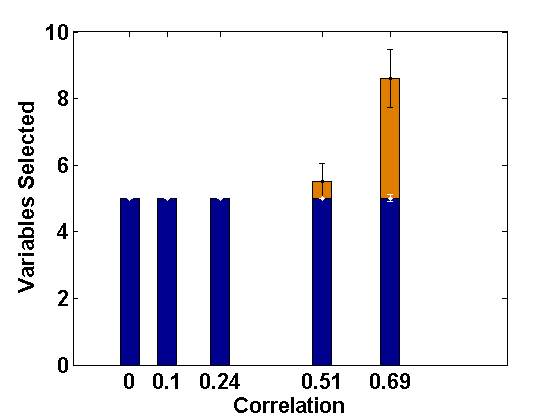}
		\label{FIG:9size_unknown_30_0}}
	~ 
	\subfloat[Endogenous variables: Set 2]{
		\includegraphics[width=0.45\textwidth]{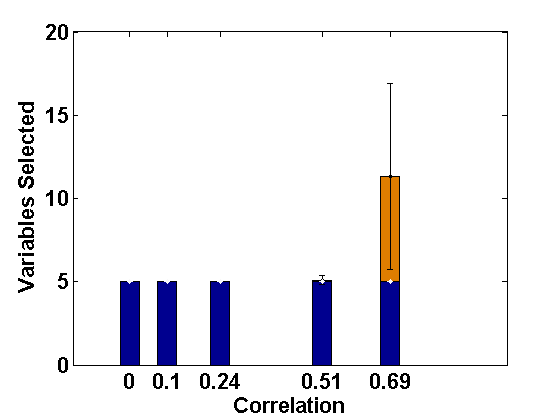}
		\label{FIG:9size_unknown_30_05}}
	\\
	\subfloat[Endogenous variables: Set 3 ]{
		\includegraphics[width=0.45\textwidth]{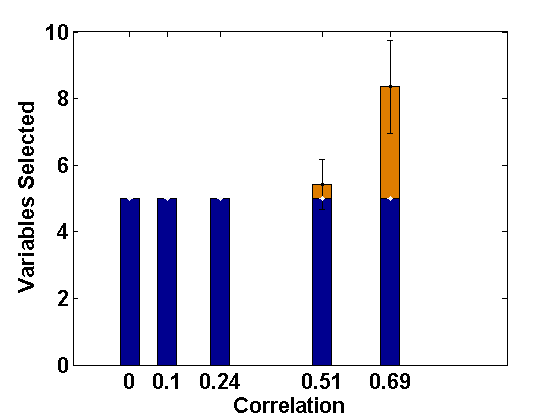}
		\label{FIG:9size_unknown_50_0}}
	~ 
	\subfloat[Endogenous variables: Set 4]{
		\includegraphics[width=0.45\textwidth]{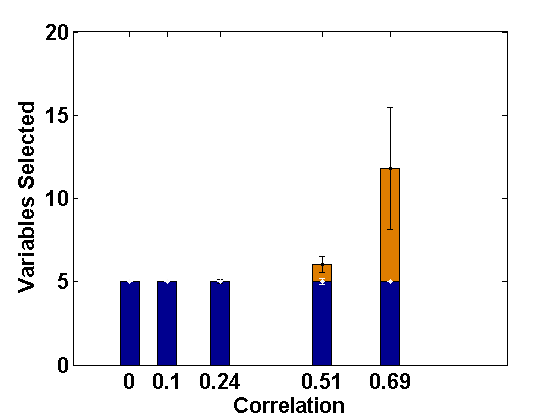}
		\label{FIG:9size_unknown_50_05}}
	\caption{Sizes of the estimated active set by the PFGMM (blue) method and the PLS method (blue+orange)
		for different extents of endogeneity and independent covariates 
		(the standard errors are shown by respective error bars)}
	\label{FIG:ActiveSet_X00}
\end{figure}

We consider the same simulation set-up as in Example \ref{SEC:MPLE}.1 with level-1 endogeneity 
and different values of the underlying parameters and apply the proposed PFGMM algorithm 
to select the relevant fixed-effects variables. 
In particular, we consider the true values of $\boldsymbol{\beta}$ as 
$\boldsymbol{\beta}=(1,2,4,3,3, 0, \ldots, 0)^T$ representing strong signal,  and  
the values of other parameters as $\sigma^2=0.25$ and $\theta_1^2=\theta_2^2=0.56$ as in Example 2.1. 
Further, we consider two values of $\rho$;  0 and $0.5$, 
indicating uncorrelated and correlated covariates, respectively, 
and different values of $\rho_e\in \{0, 0.2, 0.5, 1.5, 6\}$ to represent varying
strength of endogeneity (correlations being 0, 0.1, 0.24, 0.51, 0.69, respectively); 
note that $\rho_{e}=0$ gives the ideal case with no endogeneity.   
We have also studied negative values of $\rho_e$ with the same magnitudes leading to negative correlations,
but their effects are the same as the positive cases (only depends on the magnitudes) 
and hence the results are not reported in the paper for brevity.
The values of the regularization parameter is taken as $\lambda=0.1$, following the suggestion of Fan and Liao (2014).
For comparison, we also apply the profile likelihood proposal (referred to here as the PLS) of Fan and Li (2012).
The average sizes of the estimated active sets obtained by both methods are presented in 
Figures \ref{FIG:ActiveSet_X05} and \ref{FIG:ActiveSet_X00}, respectively, 
for the correlated and the independent covariate cases.

These empirical illustrations clearly show the significantly improved 
performance of the proposed PFGMM method under level-1 endogeneity.
In particular, for correlated covariates, even a small amount of endogeneity (small $\rho_e$)
increases the estimated active set sizes in the PLS method, 
which becomes further damaging for larger extent of endogeneity, either through more endogenous variables
or higher values of $\rho_e$. On the other hand, the proposed PFGMM method 
produces an active set of size almost equal to the original active set size (5) under any extent of endogeneity
and the variation over different replications is also negligible compared to the PLS method. 
The results for the independent variables are also similar although the harmful effect of endogeneity
on the PLS method is not significant for smaller values of $\rho_e$. 
The proposed PFGMM method still performs better than the PLS method overall, 
producing the same sets of active variables in the cases where the PLS also performs well.

The next section describes the performance of estimated regression coefficients
obtained through PFGMM, PLS and further refinements along with the estimation of the variance components.

\section{Estimation of the Variance Parameters}
\label{SEC:Var_Est}

Once we have selected the important fixed effect variables consistently through the proposed PFGMM algorithm, 
our problem reduces to a low dimensional one. Let $\widehat{S}$ denote the set of indices of 
estimated non-zero coefficients, which is asymptotically the same as the true active set $S$ 
with probability tending to one from Theorem \ref{THM:main}. 
So, we now have the reduced model
\begin{equation}
\boldsymbol{y}_i=\boldsymbol{X}_{i\widehat{S}}\boldsymbol{\beta}_{\widehat{S}} + \boldsymbol{Z}_i\boldsymbol{b}_i + \boldsymbol{\epsilon}_i,~~~~~i=1,...,I.
\label{EQ:model1_reduced}
\end{equation}

\noindent
Also, we have an estimate $\widehat{\boldsymbol{\beta}}_S$ of $\boldsymbol{\beta}_S$, 
which is consistent and asymptotically normal from Theorem \ref{THM:main}.
Then, the most straightforward and intuitive  estimates of $\boldsymbol{\eta}$
can be obtained by applying the maximum likelihood method 
to the resulting residual (random effect) model 
\begin{equation}
\widehat{\boldsymbol{r}}_i := \boldsymbol{y}_i-\boldsymbol{X}_{i\widehat{S}}\widehat{{\beta}}_S = \boldsymbol{Z}_i\boldsymbol{b}_i + \boldsymbol{\epsilon}_i,~~~~~i=1,...,I.
\label{EQ:model1_residual}
\end{equation}

\noindent
We will refer to the resulting estimator, say $\widehat{\boldsymbol{\eta}}^\ast$ 
as the PFGMM estimator of $\boldsymbol{\eta}$
in the line of the associated PFGMM estimator $\widehat{\boldsymbol{\beta}}_S$ of $\boldsymbol{\beta}_S$.
It is important to note that the PFGMME of $\boldsymbol{\eta}$ will also be consistent and asymptotically normal
by standard results on likelihood based inference for the low-dimensional residual model (\ref{EQ:model1_residual}).
Once $\widehat{\boldsymbol{\beta}}_S$ has been computed as described in Section \ref{SEC:PFGMME},
the PFGMME of $\boldsymbol{\eta}$ can be computed routinely by using the available software packages for low-dimensional LMM
(e.g., package `\textit{lme4}' in R, function `\textit{fitlme}' in MATLAB).

Alternatively, if we just want to use the proposed PFGMM for selection of important fixed effects,
in the second stage we can also fine-tune the estimates of $\boldsymbol{\beta}_{\widehat{S}}$ 
along with estimation of $\boldsymbol{\eta}$ to achieve better finite sample efficiency.
For this purpose, we consider the reduced low-dimensional linear mixed effect model given in (\ref{EQ:model1_reduced}),
containing only the $|\widehat{S}|$ fixed-effect variables from $\widehat{S}$ selected by the PFGMM algorithm,
and apply the standard maximum likelihood (ML) or the restricted maximum likelihood (REML) approach 
to get the new estimates $\widehat{\boldsymbol{\beta}}_{\widehat{S}}$ of $\boldsymbol{\beta}_{\widehat{S}}$ 
and $\widehat{\boldsymbol{\eta}}$ of $\boldsymbol{\eta}$.
Let us refer to the resulting estimators 
$((\widehat{\boldsymbol{\beta}}_{\widehat{S}}, \boldsymbol{0}), \widehat{\boldsymbol{\eta}})$ 
of $(\boldsymbol{\beta}, \boldsymbol{\eta})$ obtained by the second stage ML or REML, 
respectively, as the 2MLE or 2REMLE. 
Their performances in comparison to the PFGMM estimator of $(\boldsymbol{\beta}, \boldsymbol{\eta})$ 
are  illustrated below through a simulation.


\bigskip\noindent
\textbf{Example \ref{SEC:Var_Est}.1.} Let us repeat the simulation exercise from Section \ref{SEC:EX_PFGMM},
but now we estimate the parameters $(\boldsymbol{\beta}^T, \sigma^2, \theta_1^2, \theta_2^2)$
by the proposed PFGMM, 2MLE and 2REMLE.
The resulting mean values of the estimators, along with their standard deviation (SD) and mean squared error (MSE),
for the cases of exogeneity ($\rho_e=0$) and 
extreme endogeneity with $\rho_e=6$ are reported in Tables  \ref{TAB:Sim_Est0} and \ref{TAB:Sim_Est6}, respectively.
For comparison, we have also reported the estimates obtained by the PLS method,
where the variance parameters are estimated by maximizing the likelihood of the corresponding residual model.  

\begin{table}[!th]
	\centering
	\caption{Empirical mean, SD and MSE of different parameter estimates under no endogeneity}
	\resizebox{.8\textwidth}{!}{
		\begin{tabular}{ll|cccccc|ccc}\hline
			Method & &	$\beta_1$	&	$\beta_2$	&	$\beta_3$	&	$\beta_4$	&	$\beta_5$	&	$\beta_N$	&	
			$\theta_1^2$	&	$\theta_2^2$ & $\sigma^2$	\\	\hline
			PLS	&	Mean	&	0.988	&	1.984	&	4.030	&	3.002	&	2.985	&	0.000	&	0.544	&	0.550	&	0.230	\\
			&	SD	&	0.162	&	0.155	&	0.076	&	0.062	&	0.052	&	0.003	&	0.163	&	0.170	&	0.034	\\
			&	MSE	&	0.026	&	0.024	&	0.007	&	0.004	&	0.003	&	0.000	&	0.026	&	0.029	&	0.002	\\\hline
			PFGMM	&	Mean	&	0.987	&	1.997	&	4.003	&	2.995	&	2.996	&	0.010	&	0.544	&	0.550	&	0.230	\\
			&	SD	&	0.168	&	0.153	&	0.071	&	0.060	&	0.048	&	0.000	&	0.163	&	0.170	&	0.034	\\
			&	MSE	&	0.028	&	0.023	&	0.005	&	0.004	&	0.002	&	0.000	&	0.026	&	0.029	&	0.002	\\\hline
			2MLE	&	Mean	&	0.990	&	1.994	&	4.003	&	2.995	&	2.996	&	0.000	&	0.540	&	0.549	&	0.241	\\
			&	SD	&	0.161	&	0.153	&	0.070	&	0.059	&	0.049	&	0.000	&	0.163	&	0.173	&	0.035	\\
			&	MSE	&	0.026	&	0.023	&	0.005	&	0.004	&	0.002	&	0.000	&	0.027	&	0.030	&	0.001	\\\hline
			2REML	&	Mean	&	0.990	&	1.994	&	4.003	&	2.995	&	2.996	&	0.000	&	0.565	&	0.575	&	0.248	\\
			&	SD	&	0.161	&	0.153	&	0.070	&	0.059	&	0.049	&	0.000	&	0.170	&	0.180	&	0.036	\\
			&	MSE	&	0.026	&	0.023	&	0.005	&	0.004	&	0.002	&	0.000	&	0.028	&	0.032	&	0.001	\\\hline
		\end{tabular}
	}
	\label{TAB:Sim_Est0}
\end{table}

\begin{table}
	\centering
	\caption{Empirical mean, SD and MSE of different parameter estimates under high level-1 endogeneity with $\rho_e=6$
		and correlated covariates}
	\resizebox{.7\textwidth}{!}{
		\begin{tabular}{ll|cccccc|ccc}\hline
			Method & &	$\beta_1$	&	$\beta_2$	&	$\beta_3$	&	$\beta_4$	&	$\beta_5$	&	$\beta_N$	&	
			$\theta_1^2$	&	$\theta_2^2$ & $\sigma^2$	\\	\hline\hline
			\multicolumn{11}{l}{Endogeneity Variable: Set 1}		\\	\hline
			PLS	&	Mean	&	0.877	&	1.982	&	4.020	&	3.015	&	2.971	&	0.000	&	0.545	&	0.539	&	0.052	\\	
			&	SD	&	0.162	&	0.152	&	0.034	&	0.028	&	0.033	&	0.002	&	0.163	&	0.151	&	0.011	\\	
			&	MSE	&	0.041	&	0.023	&	0.002	&	0.001	&	0.002	&	0.000	&	0.027	&	0.023	&	0.039	\\	\hline
			PFGMM	&	Mean	&	0.965	&	1.974	&	4.004	&	2.996	&	2.994	&	0.010	&	0.545	&	0.539	&	0.052	\\	
			&	SD	&	0.212	&	0.254	&	0.069	&	0.060	&	0.050	&	0.000	&	0.163	&	0.151	&	0.011	\\	
			&	MSE	&	0.046	&	0.065	&	0.005	&	0.004	&	0.003	&	0.000	&	0.027	&	0.023	&	0.039	\\	\hline
			2MLE	&	Mean	&	0.992	&	1.992	&	4.003	&	2.996	&	2.994	&	0.000	&	0.535	&	0.548	&	0.240	\\	
			&	SD	&	0.164	&	0.155	&	0.070	&	0.059	&	0.050	&	0.000	&	0.163	&	0.173	&	0.038	\\	
			&	MSE	&	0.027	&	0.024	&	0.005	&	0.003	&	0.003	&	0.000	&	0.027	&	0.030	&	0.001	\\	\hline
			2REML	&	Mean	&	0.992	&	1.992	&	4.003	&	2.996	&	2.994	&	0.000	&	0.560	&	0.573	&	0.247	\\	
			&	SD	&	0.164	&	0.155	&	0.070	&	0.059	&	0.050	&	0.000	&	0.170	&	0.180	&	0.039	\\	
			&	MSE	&	0.027	&	0.024	&	0.005	&	0.003	&	0.003	&	0.000	&	0.028	&	0.032	&	0.001	\\	
			\hline\hline
			\multicolumn{11}{l}{Endogeneity Variable: Set 2}		\\	\hline
			PLS	&	Mean	&	0.873	&	1.979	&	4.034	&	2.990	&	3.032	&	0.000	&	0.546	&	0.540	&	0.047	\\	
			&	SD	&	0.154	&	0.150	&	0.048	&	0.033	&	0.014	&	0.004	&	0.159	&	0.152	&	0.011	\\	
			&	MSE	&	0.040	&	0.023	&	0.003	&	0.001	&	0.001	&	0.000	&	0.025	&	0.023	&	0.041	\\	\hline
			PFGMM	&	Mean	&	0.858	&	1.983	&	3.996	&	2.959	&	3.083	&	0.010	&	0.546	&	0.540	&	0.047	\\	
			&	SD	&	0.235	&	0.246	&	0.048	&	0.044	&	0.010	&	0.000	&	0.159	&	0.152	&	0.011	\\	
			&	MSE	&	0.075	&	0.060	&	0.002	&	0.004	&	0.007	&	0.000	&	0.025	&	0.023	&	0.041	\\	\hline
			2MLE	&	Mean	&	0.915	&	1.998	&	3.996	&	2.959	&	3.083	&	0.000	&	0.544	&	0.549	&	0.125	\\	
			&	SD	&	0.156	&	0.151	&	0.047	&	0.043	&	0.010	&	0.000	&	0.159	&	0.159	&	0.033	\\	
			&	MSE	&	0.031	&	0.022	&	0.002	&	0.004	&	0.007	&	0.000	&	0.025	&	0.025	&	0.017	\\	\hline
			2REML	&	Mean	&	0.915	&	1.998	&	3.996	&	2.958	&	3.083	&	0.000	&	0.568	&	0.573	&	0.129	\\	
			&	SD	&	0.156	&	0.151	&	0.047	&	0.043	&	0.010	&	0.000	&	0.166	&	0.166	&	0.033	\\	
			&	MSE	&	0.031	&	0.022	&	0.002	&	0.004	&	0.007	&	0.000	&	0.027	&	0.027	&	0.016	\\	
			\hline\hline
			\multicolumn{11}{l}{Endogeneity Variable: Set 3}		\\	\hline
			PLS	&	Mean	&	0.868	&	2.046	&	4.014	&	3.014	&	2.979	&	0.000	&	0.549	&	0.570	&	0.041	\\	
			&	SD	&	0.159	&	0.014	&	0.029	&	0.028	&	0.029	&	0.002	&	0.164	&	0.150	&	0.010	\\	
			&	MSE	&	0.042	&	0.002	&	0.001	&	0.001	&	0.001	&	0.000	&	0.027	&	0.022	&	0.044	\\	\hline
			PFGMM	&	Mean	&	0.894	&	2.094	&	3.998	&	3.007	&	2.997	&	0.010	&	0.549	&	0.570	&	0.041	\\	
			&	SD	&	0.184	&	0.009	&	0.045	&	0.043	&	0.036	&	0.000	&	0.164	&	0.150	&	0.010	\\	
			&	MSE	&	0.045	&	0.009	&	0.002	&	0.002	&	0.001	&	0.000	&	0.027	&	0.022	&	0.044	\\	\hline
			2MLE	&	Mean	&	0.905	&	2.094	&	3.998	&	3.007	&	2.996	&	0.000	&	0.548	&	0.594	&	0.105	\\	
			&	SD	&	0.158	&	0.009	&	0.045	&	0.042	&	0.036	&	0.000	&	0.164	&	0.165	&	0.026	\\	
			&	MSE	&	0.034	&	0.009	&	0.002	&	0.002	&	0.001	&	0.000	&	0.027	&	0.028	&	0.022	\\	\hline
			2REML	&	Mean	&	0.905	&	2.094	&	3.998	&	3.007	&	2.996	&	0.000	&	0.572	&	0.595	&	0.109	\\	
			&	SD	&	0.158	&	0.009	&	0.045	&	0.042	&	0.036	&	0.000	&	0.170	&	0.165	&	0.027	\\	
			&	MSE	&	0.034	&	0.009	&	0.002	&	0.002	&	0.001	&	0.000	&	0.029	&	0.028	&	0.021	\\	\hline
			\hline
			\multicolumn{11}{l}{Endogeneity Variable: Set 4}		\\	\hline
			PLS	&	Mean	&	0.842	&	1.979	&	4.034	&	3.018	&	2.983	&	0.001	&	0.538	&	0.518	&	0.005	\\	
			&	SD	&	0.156	&	0.148	&	0.019	&	0.019	&	0.017	&	0.002	&	0.158	&	0.138	&	0.002	\\	
			&	MSE	&	0.049	&	0.022	&	0.002	&	0.001	&	0.001	&	0.000	&	0.025	&	0.021	&	0.060	\\	\hline
			PFGMM	&	Mean	&	0.942	&	1.995	&	4.004	&	2.996	&	2.995	&	0.010	&	0.538	&	0.518	&	0.005	\\	
			&	SD	&	0.265	&	0.155	&	0.070	&	0.060	&	0.050	&	0.000	&	0.158	&	0.138	&	0.002	\\	
			&	MSE	&	0.073	&	0.024	&	0.005	&	0.004	&	0.002	&	0.000	&	0.025	&	0.021	&	0.060	\\	\hline
			2MLE	&	Mean	&	0.997	&	1.993	&	4.004	&	2.996	&	2.994	&	0.000	&	0.537	&	0.546	&	0.241	\\	
			&	SD	&	0.160	&	0.155	&	0.070	&	0.060	&	0.050	&	0.000	&	0.163	&	0.172	&	0.035	\\	
			&	MSE	&	0.025	&	0.024	&	0.005	&	0.004	&	0.003	&	0.000	&	0.027	&	0.030	&	0.001	\\	\hline
			2REML	&	Mean	&	0.997	&	1.993	&	4.004	&	2.996	&	2.994	&	0.000	&	0.562	&	0.572	&	0.248	\\	
			&	SD	&	0.160	&	0.155	&	0.070	&	0.060	&	0.050	&	0.000	&	0.170	&	0.180	&	0.036	\\	
			&	MSE	&	0.025	&	0.024	&	0.005	&	0.004	&	0.003	&	0.000	&	0.028	&	0.032	&	0.001	\\	\hline
		\end{tabular}
	}
	\label{TAB:Sim_Est6}
\end{table}

One can clearly observe from Table \ref{TAB:Sim_Est0} that, under exogeneity, 
the parameter estimates obtained from either of the methods are quite similar
although the estimates of error variance are slightly better through the 2MLE or 2REML approach. 
On the other hand, under endogeneity (Table \ref{TAB:Sim_Est6}), the PLS approach produces biased estimates of fixed-effect intercepts,
with larger variance and it significantly underestimates the error variance $\sigma^2$. 
The estimates obtained by the PFGMM method correct the bias of the intercept significantly
but still have somewhat larger variance of this estimate and also an underestimated value of $\sigma^2$.
However, the second stage proposal of 2MLE or 2REMLE produces highly efficient estimators 
of both the fixed-effect coefficients and variance parameters, 
which are similar to those obtained in the case of an exogenous model, 
even in the presence of extreme endogeneity of correlation $0.68$ for Sets 1, 2 and 4.
Only for Set 3, where a random slope is correlated with the error vector,
our proposed methods still have some significant (negative) bias in estimating the fixed intercept 
and error variance $\sigma^2$, although other parameters are estimated with excellent accuracy through 2MLE or 2REMLE.
One should notice that, for the two stage proposals, we are now again in a situation with endogeneity. 
See further comments related to this under Remark \ref{REM:End_slope}.

The other values of $\rho_e$ give similar results, except for Set 3, and hence they are omitted for brevity. 
In the case of endogenous random slope variables (Set 3) with moderate values of $\rho_e$ (and hence correlations)
our method is surprisingly underestimating the fixed-effect intercept term to a larger magnitude and needs further investigation;
see Remark \ref{REM:End_slope} below.

In summary, the proposed PFGMM method selects the true positive fixed-effect variables 
with extremely small amount of false positives under any extent of level-1 endogeneity, 
but the resulting estimates of fixed-effect coefficients are somewhat biased
and the resulting residual model also  underestimates the variance parameters,
specially $\sigma^2$.
However, the second stage estimators 2MLE or 2REML again correct them
to yield accurate estimators of all the parameters  under most level-1 endogeneity
except when the random slop is endogenous.

\begin{remark}[When endogeneous covariate also has random effect]\label{REM:End_slope}~~\\
	As we have already noted, although providing extremely good results in terms of our main target of fixed-effect selection, 
	the proposed FGMM as well as its second stage refinement cannot fully address the parameter estimation problem (just like PLS)
	in cases where the covariates having random effects are endogenous with the error terms. 
	However, since the proposed FGMM can select the true active sets quite accurately, 
	we can concentrate on the reduced low-dimensional model (with only the selected fixed-effect covariates) 
	to get a corrected parameter estimate in the second stage by using a suitably modified approach instead of 2MLE and 2REML. 
	Since there are already enough literature on the endogeneity issue of mixed-effects models, 
	a proper (low-dimensional) IV method, e.g., 2-stage or 3-stage least squares,  can be chosen for the above purpose 
	for a second stage refinement to PFGMM. 
	To keep the focus of the present paper clear on the fixed-effects selection in the high-dimensional context,
	we have not discussed these low-dimensional modifications for parameter estimation in the reduced model here,
	as they can be easily covered by existing literature.
	
	Another important phenomenon has been observed in our simulations with Set 3 endogenous covariates 
	and for different values of $\rho_{e}$.
	Surprisingly, the bias of the fixed effect intercept and random effect variances decreases with increasing extent of endogeneity,
	contrary to all other  cases and our standard intuition. This contradictory behavior of all the methods, PLS, PFGMM, 
	2MLE and  2REML, indicates the need for further investigation 
	and we hope to study this aspect in our future work.  
\end{remark}



\section{What happens in the Presence of Additional Level-2 endogeneity?}
\label{SEC:level2}

Although we have developed our proposed method for consistent selection of fixed effect variables 
in the LMM with level-1 endogeneity, 
it is also of interest to examine how our proposed PFGMM and its second stage refinements perform in presence of level-2 endogeneity.

\begin{figure}[!h]
	\centering
	\subfloat[Endogenous variables: Set 1]{
		\includegraphics[width=0.45\textwidth]{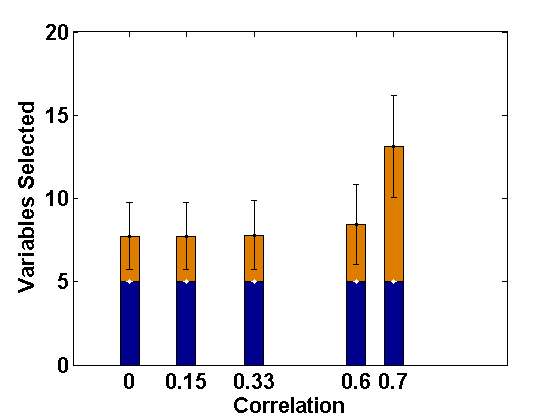}
		\label{FIG:9size_unknown_30_0}}
	~ 
	\subfloat[Endogenous variables: Set 2]{
		\includegraphics[width=0.45\textwidth]{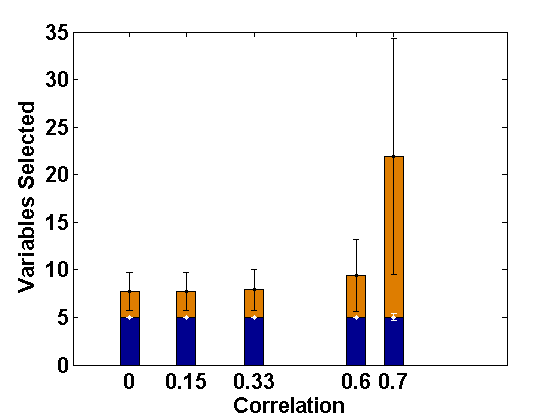}
		\label{FIG:9size_unknown_30_05}}
	\\
	\subfloat[Endogenous variables: Set 3 ]{
		\includegraphics[width=0.45\textwidth]{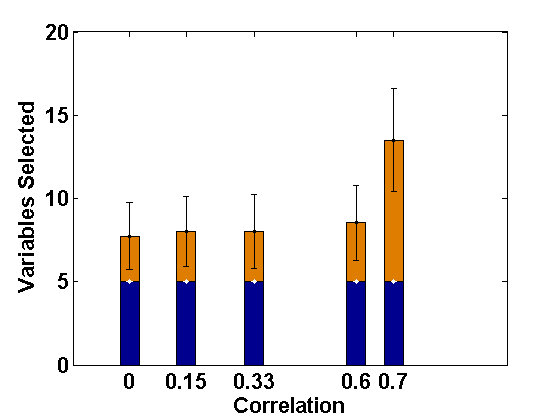}
		\label{FIG:9size_unknown_50_0}}
	~ 
	\subfloat[Endogenous variables: Set 4]{
		\includegraphics[width=0.45\textwidth]{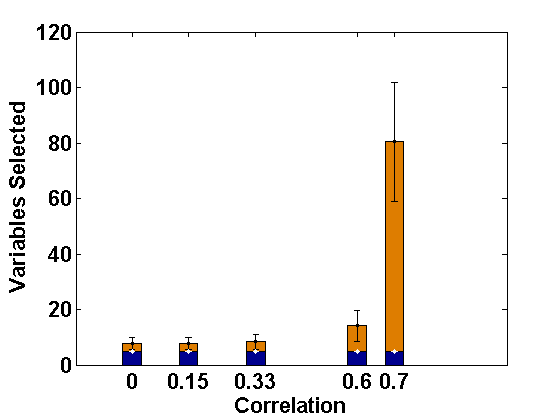}
		\label{FIG:9size_unknown_50_05}}
	\caption{Sizes of the estimated active set by the PFGMM (blue) method and the PLS method (blue+orange)
		for correlated covariates  and different extents of level-2 endogeneity in random intercept
		(the standard errors are shown by respective error bars)}
	\label{FIG:ActiveSet_RI_X05}
\end{figure}

\begin{figure}[!h]
	\centering
	\subfloat[Endogenous variables: Set 1]{
		\includegraphics[width=0.45\textwidth]{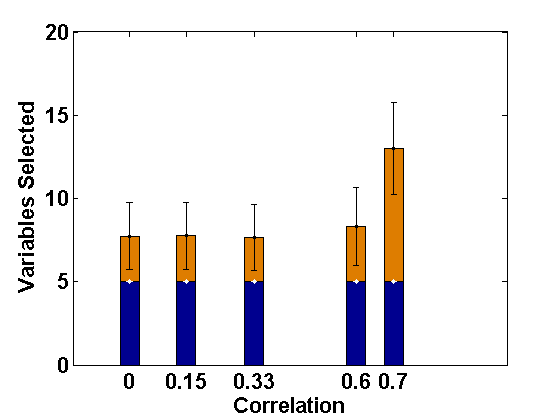}
		\label{FIG:9size_unknown_30_0}}
	~ 
	\subfloat[Endogenous variables: Set 2]{
		\includegraphics[width=0.45\textwidth]{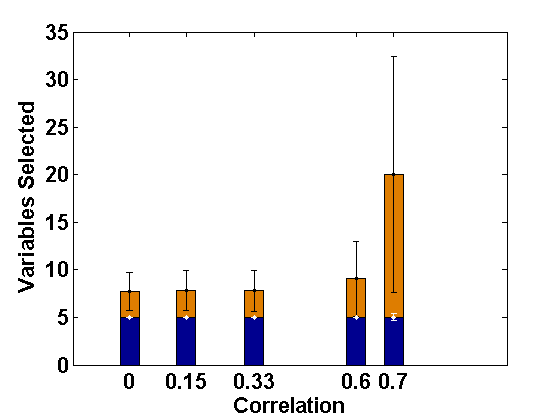}
		\label{FIG:9size_unknown_30_05}}
	\\
	\subfloat[Endogenous variables: Set 3 ]{
		\includegraphics[width=0.45\textwidth]{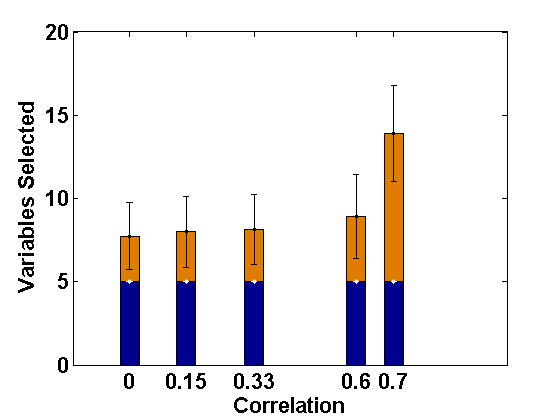}
		\label{FIG:9size_unknown_50_0}}
	~ 
	\subfloat[Endogenous variables: Set 4]{
		\includegraphics[width=0.45\textwidth]{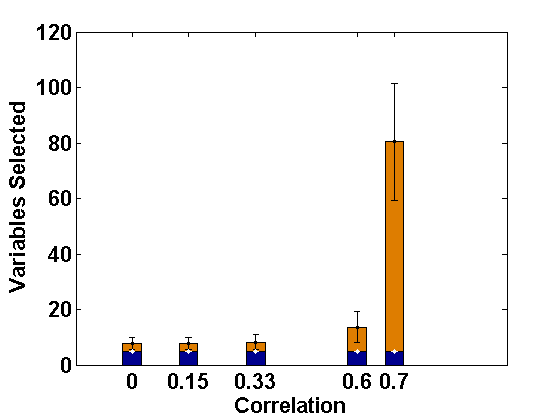}
		\label{FIG:9size_unknown_50_05}}
	\caption{Sizes of the estimated active set by the PFGMM (blue) method and the PLS method (blue+orange)
		for correlated covariates  and different extents of level-2 endogeneity in random intercept
		(the standard errors are shown by respective error bars)}
	\label{FIG:ActiveSet_RS_X05}
\end{figure}

For comparative consistency, let us again reconsider the simulation set-up of Example \ref{SEC:MPLE}.1,
but now with different extents of level-2 endogeneity in both random intercept and slope components separately 
for $\rho_b\in \{0, 0.2, 0.5, 1.5, 6\}$; this leads to correlations of 0, 0.15, 0.33, 0.6, 0.7, respectively, in both cases. 
Since the effect of level-2 endogeneity has already been observed to be significant in case of correlated covariates, 
we only present the  corresponding results regarding selection of fixed effect variables (active set sizes) 
through the usual PLS method (Fan and Li, 2012) and the PFGMM method,
and for four sets of endogenous covariates as in Example \ref{SEC:MPLE}.1. 
These are shown in Figures  \ref{FIG:ActiveSet_RI_X05} and \ref{FIG:ActiveSet_RS_X05}, respectively, 
for the cases of endogenous random intercept and slopes. 
From these Figures, as well as additional simulations not reported here, 
it has been observed that our proposed PFGMM method performs extremely well in selecting exactly the truly significant variables
compared to the PLS method even in these cases of level-2 endogeneity. Except for very high level of endogeneity,
the PFGMM method  selects exactly the   true active set in most cases as under level-1 endogeneity (or exogeneity).

Thus, if the main objective is the selection of important fixed effect variables, 
the proposed PFGMM serves the purpose in presence of any sort of endogeneity in the data,
provided the signal is reasonably  strong. 
The requirement of a strong signal is related to the choice of regularization parameter $\lambda$
and is also expected from our Assumption (A) required to prove the oracle variable selection consistency of PFGMM.
As noted in Remark \ref{REM:level2}, the theoretical derivations are also justified and linked through our numerical illustrations on this aspect.

We have also studied the effect of level-2 endogeneity on parameter estimation in different proposals
and the results are seen to be quite promising except for Set 3, as in the case of level-1 endogeneity. 
Our focus being variable selection in the present paper, for brevity, we will not discuss estimation results here.


\section{A Real Data Application}
\label{SEC:data}

We are analyzing data from a randomized controlled cross-over trial in 47 subjects (Hansson et al., 2019). 
The subjects were exposed to four different meals with similar fat contents, 
and the response was serum concentration of triglycerids (TG) measured before the meal and 2, 4, and 6 hours after. 
It is well known that an elevated level of TG is associated with an increased risk of cardiovascular disease, 
and it is of interest to understand individual variation in TG response and to characterize individuals 
with an unfavorable response. In this study we will focus on lipid subclasses.
In addition to the primary expousure (meal), we have measurements of lipid subclasses in blood, taken before each meal. 
Our primary interest is if the triglyceride response to the meal (say $y$), as measured over six hours, 
depends on the level of some of the lipid subclasses (covariates $x_j$s).
We will analyze this by a mixed model
\begin{eqnarray}
y = \beta_0 + \sum_{i=1}^{3} \beta_i D_{i} + \sum_{j=1}^{K} \gamma_j x_{j} + \sum_{i=1}^{3} \sum_{j=1}^{K} \delta_{ij} D_i x_{j}
+ b_0 + \sum_{i=1}^{3} b_i D_{i}  + \epsilon,
\end{eqnarray}

\noindent
where $D_i$, $i=1, 2, 3$, denote the dummy variables representing time 2, 4 and 6 hours, respectively,
and $K=162$ is the number of different available  lipid subclasses.
Here we have four random effect coefficients $b_i$, $i=0,1,2,3$, corresponding to random intercept and three time dummy variables.  
Additionally, we have $4(1+K)=652$  fixed effect coefficients $\beta_i$s, $\gamma_j$s and $\delta_{i,j}$s,
which need to be estimated from the repeated (incomplete) observations from only  47 patients.
However, we assume that only a few of the available lipid subclasses will 
influence the triglyceride response significantly,
and our goal is to identify these subclasses.

Therefore, we are in a sparse high-dimensional regime, 
and we can apply the proposed PFGMM method as an alternative to the PLS method
to select important lipid subclasses, assuming $b_i\sim N(0, \sigma_i^2)$ for $i=0,1,2,3$, and $\epsilon\sim N(0, \sigma^2)$. 
It is difficult to test for endogeneity in high-dimensional models in practice. 
However, due to the high dimensions one will almost expect endogeneity to arise incidentally. 
In the current example, important potential confounders are also omitted from the model, 
which would be expected to lead to endogeneity problems. 
From our simulation studies in Section \ref{SEC:EX_PFGMM} 
we observed that the PLS method has serious problems with over-selection in situations with endogenous covariates, 
and hence, we will consider a large reduction in the number of selected variables by our method as a sign of endogeneity.

We have applied both the methods with different values of the regularization parameter $\lambda$
for the purpose above. We observe that, in each case, the PFGMM method selects way less significant variables 
compared to the existing PLS approach, giving us less false positives; 
see Table \ref{TAB:Data_VarEst} for a few illustrative cases. 
This clearly indicates the presence of significant endogeneity in the data 
and the advantages of our proposed PFGMM approach becomes clear.

\begin{table}[h]
	\centering
	\caption{Estimated active set size $|S|$ and variance components $\sigma_i^2$ ($i=0, 1, 2,3$) and $\sigma^2$, for the real data example,
		obtained by PLS and the proposed PFGMM combined with 2REML}
	\resizebox{.8\textwidth}{!}{
		\begin{tabular}{ll|r|ccccc}\hline
			$\lambda(\times10^{-3})$ & Method & $S$ &	$\sigma^2_0$	&	$\sigma^2_1$	&	$\sigma^2_2$	&	$\sigma^2_3$ & $\sigma^2$	\\	\hline\hline
			5	&	PLS	&	39	&	0.050	&	0.057	&	0.028	&	0.045	&	4.42E-07\\
			&	PFGMM+2REML	&	5	&	0.146	&	0.126	&	0.053	&	0.073	&	9.66E-08 \\\hline
			2	&	PLS	&	113	&	0.030	&	0.048	&	0.022	&	0.039	&	7.33E-09\\
			&	PFGMM+2REML	&	11	&	0.146	&	0.098	&	0.031	&	0.063	&	2.13E-07 \\\hline
			1.5	&	PLS	&	136	&	0.026	&	0.046	&	0.020	&	0.033	&	2.77E-09\\
			&	PFGMM+2REML	&	30	&	0.046	&	0.059	&	0.053	&	0.047	&	1.50E-07 \\\hline
		\end{tabular}
	}
	\label{TAB:Data_VarEst}
\end{table}

In the same Table \ref{TAB:Data_VarEst}, the variance estimates obtained by the second stage refinement 2REML are also reported;
clearly the error variance reduces as we select more and more  fixed effect variables by lower $\lambda$-values.
The appropriate model can be chosen via proper justification along with biological significance of the resulting model estimates.
For example, the model with $\lambda=2\times10^{-3}$ selecting 11 fixed effects looks the best candidate for the present example,
since it provides a very low model error and still a rather sparse model. 
Of particular interest is the selection of a total of seven interaction parameters, 
pointing to subclasses of particular interest when it comes to triglyceride response. 
Without going into detail about the lipid subclasses, two of the discoveries seem obvious, 
as they are related to subclasses rich in triglycerides. 
Furthermore, four parameters are pointing to subclasses related to Hdl cholesterol, 
a parameter known to be connected to triglycerides, 
while the significance of the last subclass being picked up is more unclear.

As an alternative to the model selection above, one can apply 
a proper extension of BIC to chose a data driven value of the regularization parameter $\lambda$. 
Some indications are provided in Section \ref{SEC:PFGMM_Computation} 
since the usual BIC often gets affected by the presence of endogeneity in the data. 
However, more investigation is needed on appropriate BIC extensions under endogeneity 
which we hope to do in future work.

Finally, we should mention that we have used the unimportant covariates as a general vector of instrumental variables,
which is seen to perform well in all our simulations studies;
such instruments were also suggested by Fan and Liao (2014) 
while dealing with high-dimensional regression models with endogeneity. 
More detailed study on finding an optimal instrumental variable will surely 
be an interesting future research problem.

\section{Conclusions}
\label{SEC:conclusions}

In this paper we have studied the problem of endogeneity in high-dimensional LMM
with particular attention to the selection of important fixed-effect variables under error-covariate endogenity.
We have proved the inconsistency of the usual penalized likelihood approach for such cases
and proposed a new PFGMM approach of consistent selection of the fixed-effects 
combining the ideas of generalized method of moments, instrumental variables 
and proxy matrix for the unknown variance component matrix. 
The oracle variable selection property as well as the consistency 
and asymptotic normality of the estimated fixed effects coefficients are derived under appropriate assumptions.

This work opens up many different new research questions for future research. 
The immediate follow-up would be the detailed analysis of level-2 endogeneity 
and its effect on the usual likelihood method as well as our proposed PFGMM method.
We should develop appropriate modifications in such cases, if needed, to establish variable selection consistency. 
The second stage estimators may be further investigated for theoretical optimality. 
Further, a suitable extension of BIC should be studied, both theoretically and empirically, 
to select the regularization parameter from endogenous data. Although we have indicated a possible solution,
detailed analysis 
is due for future works.  

\bigskip\noindent {\bf Acknowledgements.}\\
The majority of this work was done when the first author was visiting the University of Oslo, Norway, 
through  the support of a grant from the Norwegian Cancer Society.
The research of the first author is also partially supported by the INSPIRE Faculty Research Grant 
from Department of Science and Technology, Govt. of India.

\newpage
\appendix
\section{Online Supplementary Material}
\subsection{Proof of Theorem 1}\label{App:proof1}

Define $P_{n,\lambda}'(0+)=\lim_{t\rightarrow0+} P_{n, \lambda}'(t)$.
Then, by an application of the Karush-Kuhn-Tucker (KKT) condition on the local minimizers 
$\widehat{\boldsymbol\beta}$ and $\widehat{\boldsymbol\eta}$, we get 
$$
\frac{\partial L_n(\widehat{\boldsymbol\beta},\widehat{\boldsymbol\eta})}{\partial \beta_l} + v_l = 0, 
~~~~\mbox{for all } l\leq p,
$$
where $v_i = P_{n,\lambda}(|\widehat{\beta}_l|)\mbox{sgn}(\widehat{\beta}_l)$ if $\widehat{\beta}_l\neq 0$,
and $v_i \in [-P_{n,\lambda}'(0+), P_{n,\lambda}'(0+)]$ if $\widehat{\beta}_l= 0$.
Therefore, by using the monotonicity and the limit of $P_{n, \lambda}'(t)$ from Condition (C3), we get
\begin{equation}
\left|\frac{\partial L_n(\widehat{\boldsymbol\beta},\widehat{\boldsymbol\eta})}{\partial \beta_l}\right|
\leq P_{n,\lambda}'(0+) = o(1).
\label{EQ:p1}
\end{equation}
Next, by the first order Taylor series expansion of $\frac{\partial L_n(\boldsymbol\beta,\boldsymbol\eta)}{\partial \beta_l}$
at $(\widehat{\boldsymbol{\beta}}, \widehat{\boldsymbol{\eta}})$ around $(\boldsymbol{\beta}_0, \boldsymbol{\eta}_0)$,
we get a $(\widetilde{\boldsymbol{\beta}}, \widetilde{\boldsymbol{\eta}})$ on the line segment joining 
$(\widehat{\boldsymbol{\beta}}, \widehat{\boldsymbol{\eta}})$ and $(\boldsymbol{\beta}_0, \boldsymbol{\eta}_0)$
such that 
$$
\frac{\partial L_n(\widehat{\boldsymbol\beta},\widehat{\boldsymbol\eta})}{\partial \beta_l}
- \frac{\partial L_n(\boldsymbol\beta_0,\boldsymbol\eta_0)}{\partial \beta_l}
= \sum_{j=1}^p\frac{\partial^2 L_n(\widetilde{\boldsymbol\beta},\widetilde{\boldsymbol\eta})}{\partial \beta_l\beta_j} 
(\widehat{\beta}_j - \beta_{0j}) 
+ \sum_{k=1}^m \frac{\partial^2 L_n(\widetilde{\boldsymbol\beta},\widetilde{\boldsymbol\eta})}{\partial \beta_l\eta_k}
(\widehat{\eta}_k - \eta_{0k}).
$$
Therefore, in the event $\widehat{\boldsymbol\beta}_N=0$ having probability tending to one [by Condition (C2)],
we get 
\begin{eqnarray}
&&\left|\frac{\partial L_n(\widehat{\boldsymbol\beta},\widehat{\boldsymbol\eta})}{\partial \beta_l}
- \frac{\partial L_n(\boldsymbol\beta_0,\boldsymbol\eta_0)}{\partial \beta_l}\right|
\nonumber\\
&&= \left|\sum_{j\in S}\frac{\partial^2 L_n(\widetilde{\boldsymbol\beta},\widetilde{\boldsymbol\eta})}{\partial \beta_l\beta_j} 
(\widehat{\beta}_j - \beta_{0j}) 
+ \sum_{k=1}^m \frac{\partial^2 L_n(\widetilde{\boldsymbol\beta},\widetilde{\boldsymbol\eta})}{\partial \beta_l\eta_k}
(\widehat{\eta}_k - \eta_{0k})\right|
\nonumber\\
&&
\leq  \max_{l,j\leq p}
\left|\frac{\partial^2 L_n(\widetilde{\boldsymbol\beta},\widetilde{\boldsymbol\eta})}{\partial \beta_l\beta_j}\right| 
\left|\left|\widehat{\boldsymbol{\beta}}_S - \boldsymbol{\beta}_{0S}\right|\right|_1
+  \max_{l,k\leq p}
\left|\frac{\partial^2 L_n(\widetilde{\boldsymbol\beta},\widetilde{\boldsymbol\eta})}{\partial \beta_l\eta_k}\right| 
\left|\left|\widehat{\boldsymbol{\eta}}- \boldsymbol{\eta}_{0}\right|\right|_1
\nonumber\\
&&
\leq  \max_{l,j\leq p}
\left|\frac{\partial^2 L_n(\widetilde{\boldsymbol\beta},\widetilde{\boldsymbol\eta})}{\partial \beta_l\beta_j}\right| 
\sqrt{s}\left|\left|\widehat{\boldsymbol{\beta}}_S - \boldsymbol{\beta}_{0S}\right|\right|_2
+  \max_{l,k\leq p}
\left|\frac{\partial^2 L_n(\widetilde{\boldsymbol\beta},\widetilde{\boldsymbol\eta})}{\partial \beta_l\eta_k}\right| 
\sqrt{m}\left|\left|\widehat{\boldsymbol{\eta}}- \boldsymbol{\eta}_{0}\right|\right|_2,
\nonumber
\end{eqnarray}
where the last step follows by Cauchy-Swartz inequality;
here $m$ is the dimension of $\boldsymbol\eta$. Now, by Conditions (C1) and (C2), we get 
$$
\left|\frac{\partial L_n(\widehat{\boldsymbol\beta},\widehat{\boldsymbol\eta})}{\partial \beta_l}
- \frac{\partial L_n(\boldsymbol\beta_0,\boldsymbol\eta_0)}{\partial \beta_l}\right|=o_P(1).
$$
Then the theorem follows using (\ref{EQ:p1})
\hfill{$\square$}

\subsection{Proof of Theorem 2}\label{App:proof2}
First let us note that, for the likelihood loss 
$L_n(\boldsymbol\beta_0,\boldsymbol\eta_0) = -l_n(\boldsymbol{\beta}, \boldsymbol{\eta})$, we have
$$
\frac{\partial L_n(\boldsymbol\beta_0,\boldsymbol\eta_0)}{\partial \beta_k}
= - \frac{1}{\sigma^2}(\boldsymbol{y}-\boldsymbol{X}\boldsymbol{\beta})^T
\boldsymbol{V}(\boldsymbol{\theta},\sigma^2)^{-1}\boldsymbol{X}^{(k)},
$$
for any $k\leq p$, where $\boldsymbol{X}^{(k)}$ denotes the $k$-th column of the matrix $\boldsymbol{X}$.
Therefore, by an application of Strong law of Large Numbers, we have the following result
in terms of the transformed regression model  given in Equation (3.7) of the main paper:
\begin{eqnarray}
\left|\frac{\partial L_n(\boldsymbol\beta_0,\boldsymbol\eta_0)}{\partial \beta_k}\right|
\rightarrow E(\epsilon^\ast X_k^\ast),
~~\mbox{almost surely}, ~~\mbox{as } n\rightarrow\infty,
\label{EQ:pf2}
\end{eqnarray}
where $\epsilon^\ast$ and $X_k^\ast$ represent the random variables corresponding to the transformed
error $\boldsymbol{\epsilon}^\ast = \boldsymbol{V}(\boldsymbol{\theta},\sigma^2)^{-1/2}\boldsymbol{\epsilon}$
and the $k$-th transformed covariate (column) in 
$\boldsymbol{X}^\ast = \boldsymbol{V}(\boldsymbol{\theta},\sigma^2)^{-1/2}\boldsymbol{X}$.
Now, if $X_k$ is endogenous, then clearly $\epsilon^\ast$ and $X_k^\ast$ will be correlated
and hence the limit in (\ref{EQ:pf2}) will be non-zero. 
Then the proof  follows directly from the results of Theorem 1.
\hfill{$\square$}

\subsection{Proof of Theorem 3}\label{App:proof}

We will first show that our Assumptions (A), (I) and (M) together with (P) imply the following four results 
for the PFGMM loss function $L_n^P(\boldsymbol{\beta})$ given in Eq.~3.11 of the main paper.
\begin{enumerate}
	\item[(R1)] $\left|\left|\nabla_SL_n^P(\boldsymbol{\beta}_{0S},\boldsymbol{0})\right|\right| = O_P\left(\sqrt{\frac{s\log p}{n}}\right)$,
	where $\nabla_S$ denotes the gradient with respect to the (non-zero) elements of $\boldsymbol{\beta}$ in $S$.
	Note that $\sqrt{\frac{s\log p}{n}}=o(d_n)$ by our Assumptions.
	
	\item[(R2)] For any $\epsilon>0$, there exists a positive constant $C_\epsilon$ such that, for all sufficiently large $n$,  
	$$
	P\left(\lambda_{\min}\left[\nabla_S^2 L_n^P(\boldsymbol{\beta}_{0S},\boldsymbol{0})\right]>C_\epsilon\right) > 1 - \epsilon.
	$$
	
	\item[(R3)] For any $\epsilon>0$, $\delta>0$ and any non-negative sequence $\alpha_n=o(d_n)$, 
	there exists a positive integer $N$ such that, for all $n\geq N$,  
	$$
	P\left(\sup\limits_{||\boldsymbol{\beta}_S - \boldsymbol{\beta}_{0S}||\leq \alpha_n}\left|\left|
	\nabla_S^2 L_n^P(\boldsymbol{\beta}_{S},\boldsymbol{0})- \nabla_S^2 L_n^P(\boldsymbol{\beta}_{0S},\boldsymbol{0})\right|\right|_F
	\leq \delta\right) > 1 - \epsilon,
	$$
	where $||\boldsymbol{A}||_F$ denotes the Frobenius norm of a matrix $\boldsymbol{A}$.
	
	\item[(R4)] For any $\epsilon>0$, there exists a positive constant $C_\epsilon$ such that, for all sufficiently large $n$,  
	$$
	P\left(\lambda_{\min}\left[\nabla_S^2 L_n^P(\boldsymbol{\beta}_{0S},\boldsymbol{0})\right]>C_\epsilon\right) > 1 - \epsilon.
	$$
\end{enumerate}
Then, Parts (a) and (b) of our Theorem 3 follow from Theorems B.1 and B.2 of Fan and Liao (2014).
Note that the assumptions required on the penalty functions there are exactly the same as our Assumption (P);
see Fan and Liao (2014) for details. 

In the following, we will use the notations $\boldsymbol{\Pi}_S = \boldsymbol{\Pi}(\boldsymbol{\beta}_{0S})$ and
\begin{eqnarray}
\widetilde{L_n^P}(\boldsymbol{\beta}_S) &=& 
\left[\frac{1}{n}\boldsymbol{\Pi}_S\widetilde{\boldsymbol{V}}_z^{-1} (\boldsymbol{y} - \boldsymbol{X}_S\boldsymbol{\beta}_S)\right]^T
\boldsymbol{J}(\boldsymbol{\beta}_0) \left[\frac{1}{n}\boldsymbol{\Pi}_S\widetilde{\boldsymbol{V}}_z^{-1} 
(\boldsymbol{y} - \boldsymbol{X}_S\boldsymbol{\beta}_S)\right],
~~~~ \boldsymbol{\beta}_S \in \mathbb{R}^s.
\label{EQ:PFGMM_loss1}
\end{eqnarray}

\noindent
Note that $\widetilde{L_n^P}(\boldsymbol{\beta}_S) = L_n^P(\boldsymbol{\beta}_{S},\boldsymbol{0})$. 
We will now prove results (R1)--(R4). 

\bigskip
\noindent
\textbf{Proof of (R1):}\\
By standard derivative calculations, we get 
$\nabla \widetilde{L_n^P}(\boldsymbol{\beta}_{S}) = 2 \boldsymbol{A}_n(\boldsymbol{\beta}_{S})\boldsymbol{J}(\boldsymbol{\beta}_0)
\left[\frac{1}{n}\boldsymbol{\Pi}_S\widetilde{\boldsymbol{V}}_z^{-1}(\boldsymbol{y} - \boldsymbol{X}_S\boldsymbol{\beta}_S)\right],$
where $\boldsymbol{A}_n(\boldsymbol{\beta}_S) = - \frac{1}{n} \left(\boldsymbol{\Pi}_S\widetilde{\boldsymbol{V}}_z^{-1}\boldsymbol{X}_S \right)$.
Now, by Assumption (I4), we know that $\left|\left|\boldsymbol{A}_n(\boldsymbol{\beta}_{0S})\right|\right|=O_P(1)$.
Also, by Assumption (I2), the elements in $\boldsymbol{J}(\boldsymbol{\beta}_0)$ are uniformly bounded in probability, 
and hence 
\begin{eqnarray}
\left|\left|\nabla \widetilde{L_n^P}(\boldsymbol{\beta}_{0S})\right|\right| \leq O_P(1) 
\left|\left|\frac{1}{n}\boldsymbol{\Pi}_S\widetilde{\boldsymbol{V}}_z^{-1}(\boldsymbol{y} - \boldsymbol{X}_S\boldsymbol{\beta}_{0S})\right|\right|.
\label{EQ:C1_Eq0}
\end{eqnarray}

\noindent
Next, we study the difference of the random variables 
$\boldsymbol{Z}_1 = \left[\frac{1}{n}\boldsymbol{\Pi}_S\widetilde{\boldsymbol{V}}_z^{-1}(\boldsymbol{y} - \boldsymbol{X}_S\boldsymbol{\beta}_{0S})\right]$ 
and 
$\boldsymbol{Z}_2=\left[\frac{1}{n}\boldsymbol{\Pi}_S\boldsymbol{V}(\boldsymbol{\theta})^{-1}(\boldsymbol{y} - \boldsymbol{X}_S\boldsymbol{\beta}_{0S})\right]$.
By Assumption (M1), we get 
$$
C_1 \widetilde{\boldsymbol{V}}_z - \boldsymbol{V}(\boldsymbol{\theta}, \sigma^2) = (C_1 - 1)\boldsymbol{I} 
+ \boldsymbol{Z}^T(C_1\mathcal{M} - \sigma^{-2}\boldsymbol{\Psi}_{\boldsymbol{\theta}})\boldsymbol{Z} \geq 0.
$$ 
That is $C_1 \widetilde{\boldsymbol{V}}_z \geq  \boldsymbol{V}(\boldsymbol{\theta}, \sigma^2)$. By the Woodbury formula, 
since $C_1 \widetilde{\boldsymbol{V}}_z$ and $\boldsymbol{V}(\boldsymbol{\theta}, \sigma^2)$ are both positive definite, we get 
$\widetilde{\boldsymbol{V}}_z^{-1} \leq C_1 \boldsymbol{V}(\boldsymbol{\theta}, \sigma^2)^{-1}$. Therefore, 
\begin{eqnarray}
\boldsymbol{Z}_1 - \boldsymbol{Z}_2 &=& 
\frac{1}{n} \boldsymbol{\Pi}_S \left[\widetilde{\boldsymbol{V}}_z^{-1} - \boldsymbol{V}(\boldsymbol{\theta}, \sigma^2)^{-1}\right]
(\boldsymbol{y} - \boldsymbol{X}_S\boldsymbol{\beta}_{0S})
\nonumber\\
&\leq& \frac{(C_1-1)}{n} \boldsymbol{\Pi}_S \boldsymbol{V}(\boldsymbol{\theta}, \sigma^2)^{-1}(\boldsymbol{y} - \boldsymbol{X}_S\boldsymbol{\beta}_{0S}).
\label{EQ:C1_Eq1}
\end{eqnarray}

\noindent
Further, by Assumption (M2), we have 
$$
C_1 (\log n)\boldsymbol{V}(\boldsymbol{\theta}, \sigma^2) - \widetilde{\boldsymbol{V}}_z 
= (C_1\log n - 1)\boldsymbol{I} + \boldsymbol{Z}^T(C_1\log n\sigma^{-2}\boldsymbol{\Psi}_{\boldsymbol{\theta}}- \mathcal{M})\boldsymbol{Z} \geq 0.
$$ 
Then, $C_1 (\log n)\boldsymbol{V}(\boldsymbol{\theta}, \sigma^2)\geq \widetilde{\boldsymbol{V}}_z$, and as before we get 
$C_1(\log n)\widetilde{\boldsymbol{V}}_z^{-1} \geq \boldsymbol{V}(\boldsymbol{\theta}, \sigma^2)^{-1}$. Therefore, 
\begin{eqnarray}
\boldsymbol{Z}_2 - \boldsymbol{Z}_1 &=& 
\frac{1}{n} \boldsymbol{\Pi}_S \left[\boldsymbol{V}(\boldsymbol{\theta}, \sigma^2)^{-1} - \widetilde{\boldsymbol{V}}_z^{-1}\right]
(\boldsymbol{y} - \boldsymbol{X}_S\boldsymbol{\beta}_{0S})
\nonumber\\&\leq& 
\leq \frac{C_1(C_1\log n -1)}{n} 
\boldsymbol{\Pi}_S \boldsymbol{V}(\boldsymbol{\theta}, \sigma^2)^{-1}(\boldsymbol{y} - \boldsymbol{X}_S\boldsymbol{\beta}_{0S}).
\label{EQ:C1_Eq2}
\end{eqnarray}

\noindent
Combining (\ref{EQ:C1_Eq1}) and (\ref{EQ:C1_Eq2}), along with our basic IV assumption (Eq. (3.8) of the main paper), 
we have $\left|\boldsymbol{Z}_1 - \boldsymbol{Z}_2\right| = o_P(1)$.
Therefore, from (\ref{EQ:C1_Eq0}), we get
\begin{eqnarray}
\left|\left|\nabla \widetilde{L_n^P}(\boldsymbol{\beta}_{0S})\right|\right| 
&\leq& O_P(1)\left|\left|\frac{1}{n}\boldsymbol{\Pi}_S\boldsymbol{V}(\boldsymbol{\theta}, \sigma^2)^{-1}
(\boldsymbol{y} - \boldsymbol{X}_S\boldsymbol{\beta}_{0S})\right|\right|
\nonumber\\&=& 
O_P(1)\left|\left|\frac{1}{n}\sum_{i=1}^n ({y}_i^\ast - \boldsymbol{X}_{iS}^\ast\boldsymbol{\beta}_{0S})\Pi_{iS}^\ast\right|\right|.~~~~
\label{EQ:C1_Eq00}
\end{eqnarray}

\noindent
But, $E\left[(Y^\ast - \boldsymbol{X}_i^\ast\boldsymbol{\beta}_{0S})\Pi_i^\ast\right]=0$ by the choice of IV $\pi_i^\ast$.
So, using the Bonferroni inequality and the exponential-tail Bernstein inequality along with Assumption (I1) 
and the normality of $(Y^\ast - \boldsymbol{X}_i^\ast\boldsymbol{\beta}_{0S})$, 
we get a positive constant $C$ such that, for any $t>0$, 
\begin{eqnarray}
P\left(\max_{l\leq p}\left|\frac{1}{n}\sum_{i=1}^n ({y}_i^\ast - \boldsymbol{X}_{iS}^\ast\boldsymbol{\beta}_{0S})F_{li}^\ast\right|>t\right)
&<& p \max_{l\leq p} P\left(\left|\frac{1}{n}\sum_{i=1}^n ({y}_i^\ast - \boldsymbol{X}_{iS}^\ast\boldsymbol{\beta}_{0S})F_{li}^\ast\right|>t\right)
\nonumber\\&\leq& 
\leq p \exp(- {Ct^2}/{n}).\nonumber
\end{eqnarray}
Thus, 
$$
P\left(\max_{l\leq p}\left|\frac{1}{n}\sum_{i=1}^n ({y}_i^\ast - \boldsymbol{X}_{iS}^\ast\boldsymbol{\beta}_{0S})F_{li}^\ast\right|>t\right)
=O_P\left(\sqrt{\frac{\log p}{n}}\right).
$$ 
Similarly, we can show
$$
P\left(\max_{l\leq p}\left|\frac{1}{n}\sum_{i=1}^n ({y}_i^\ast - \boldsymbol{X}_{iS}^\ast\boldsymbol{\beta}_{0S})H_{li}^\ast\right|>t\right)
=O_P\left(\sqrt{\frac{\log p}{n}}\right).
$$ 
Combining with (\ref{EQ:C1_Eq00}) we get 
$\left|\left|\nabla \widetilde{L_n^P}(\boldsymbol{\beta}_{0S})\right|\right| =O_P\left(\sqrt{\frac{s\log p}{n}}\right)$, proving (R1). 
\hfill{$\square$}

\bigskip\bigskip
\noindent
\textbf{Proof of (R2):}\\
Note that, by standard derivative calculations, we have 
$
\nabla^2 \widetilde{L_n^P}(\boldsymbol{\beta}_{0S}) = 2 \boldsymbol{A}_n(\boldsymbol{\beta}_{0S})\boldsymbol{J}(\boldsymbol{\beta}_0)
\boldsymbol{A}_n(\boldsymbol{\beta}_{0S})^T.
$

\noindent
Fix any $\epsilon>0$. By Assumption (I2), there exists a constant $C>0$ such that 
$P(\lambda_{\min}[\boldsymbol{J}(\boldsymbol{\beta}_0)]>C)>1-\epsilon$ for all sufficiently large $n$.
Also, by Assumption (I4), there exists a constant $C_2>0$ such that $\lambda_{\min}[\boldsymbol{A}\boldsymbol{A}^T] >C_2$, 
where $\boldsymbol{A}$ is as defined in Assumption (I4). Now, let us consider the events
$$
G_1 =\left\{\lambda_{\min}[\boldsymbol{J}(\boldsymbol{\beta}_0)]>C\right\},
~~~~
G_2=\left\{\left|\left|\boldsymbol{A}_n(\boldsymbol{\beta}_{0S})\boldsymbol{A}_n(\boldsymbol{\beta}_{0S})^T
- \boldsymbol{A}\boldsymbol{A}^T\right|\right|<\frac{C_2}{2}\right\}.
$$
On the event $G_1\cap G_2$, we have

\begin{eqnarray}
\lambda_{\min}\left[\nabla^2 \widetilde{L_n^P}(\boldsymbol{\beta}_{0S})\right]
&\geq& 2 \lambda_{\min}[\boldsymbol{J}(\boldsymbol{\beta}_0)]
\lambda_{\min}\left[\boldsymbol{A}_n(\boldsymbol{\beta}_{0S})\boldsymbol{A}_n(\boldsymbol{\beta}_{0S})^T\right]
\nonumber\\&\geq& 
\geq 2C\left\{\lambda_{\min}[\boldsymbol{A}\boldsymbol{A}^T]  - \frac{C_2}{2}\right\}
>CC_2.~~~~~~~~~~
\end{eqnarray}

\noindent
But, we already have $P(G_1)>1-\epsilon$. And, by the definition of matrix $\boldsymbol{A}$, we have
$P(G_2^c)<\epsilon$ for all sufficiently large $n$. Hence 
$P(G_1\cap G_2) \geq 1-P(G_1^c) - P(G_2^c)>1-2\epsilon,$ which completes the proof of (R2).
\hfill{$\square$}

\bigskip
\noindent
\textbf{Proof of (R3):}\\
Fix any $\epsilon>0$, $\delta>0$ and any non-negative sequence $\alpha_n = o(d_n)$. 
For all $\boldsymbol{\beta}_S$ satisfying $||\boldsymbol{\beta_S} - \boldsymbol{\beta}_{0S}||<d_n/2$, we have $\beta_{S,k}\neq 0$ for all $k\leq s$.
Thus, $\boldsymbol{J}(\boldsymbol{\beta}_S)=\boldsymbol{J}(\boldsymbol{\beta}_{0S})$.
Also 
$$
P\left(\sup\limits_{||\boldsymbol{\beta}_S - \boldsymbol{\beta}_{0S}||\leq \alpha_n}
\left|\left|\boldsymbol{A}_n(\boldsymbol{\beta}_{S})- \boldsymbol{A}_n(\boldsymbol{\beta}_{0S})\right|\right|_F
\leq \delta\right) > 1 - \epsilon.
$$
Combining we get	
$$
P\left(\sup\limits_{||\boldsymbol{\beta}_S - \boldsymbol{\beta}_{0S}||\leq \alpha_n}\left|\left|
\nabla_S^2 \widetilde{L_n^P}(\boldsymbol{\beta}_{S})- \nabla_S^2 \widetilde{L_n^P}(\boldsymbol{\beta}_{0S})\right|\right|_F
\leq \delta\right) > 1 - \epsilon,
$$
which completes the proof of (R3). 
\hfill{$\square$}

\bigskip
\noindent
\textbf{Proof of (R4):}\\
The proof follows in the same line of argument as in Appendix C.1.2 of Fan and Liao (2014)
and hence left out for brevity.
\hfill{$\square$}

\bigskip\bigskip
\noindent
\textbf{Proof of Parts (a)-(b) of Theorem 3:}\\
Under the results (R1)--(R4) along with Assumption (P), we can apply Theorem B.2 of Fan and Liao (2014) for our PFGMM loss 
to conclude Part (a) of Theorem 3,
and we also get that $P(\widehat{S}\subset S) \rightarrow 1$. 
Further, from Theorem B.1 of Fan and Liao (2014), we have $\left|\left|\widehat{\boldsymbol{\beta}}_S - \boldsymbol{\beta}_{0S}\right|\right|=o_P(d_n)$.
Then, 
\begin{eqnarray}
P(S\nsubseteq\widehat{S}) &=& P(\mbox{There exists a } j\in S \mbox{ such that } \widehat{{\beta}}_j=0)
\nonumber\\
&\leq& P(\mbox{There exists a } j\in S \mbox{ such that } |\widehat{{\beta}}_j-\beta_{0j}|\geq|\beta_{0j}|)
\nonumber\\
&\leq& P(\max_{j\in S}|\widehat{{\beta}}_j-\beta_{0j}|\geq d_n )
\nonumber\\
&\leq& P(||\widehat{{\beta}}_j-\beta_{0j}||\geq d_n ) =o(1).
\end{eqnarray}

\noindent
Therefore, $P(S\subset \widehat{S})\rightarrow 1$, and hence $P(\widehat{S}=S)\rightarrow 1$.
\hfill{$\square$}

\bigskip\newpage
\noindent
\textbf{Proof of Part (c) of Theorem 3:}\\
We start with the KKT condition for $\widehat{\boldsymbol{\beta}}_S$ which gives
$$
-P_n'(|\widehat{\boldsymbol{\beta}}_S|)\circ \mbox{sgn}(\widehat{\boldsymbol{\beta}}_S) 
= \nabla \widetilde{L_n^P}(\widehat{\boldsymbol{\beta}}_S),
$$
where sgn denote the sign function, $\circ$ denotes the element-wise product and 
$$
P_n'(|\widehat{\boldsymbol{\beta}}_S|)=(P_{n,\lambda}(|\widehat{{\beta}}_{S,1}|), \ldots, P_{n,\lambda}(|\widehat{{\beta}}_{S,s}|))^T.
$$
By the Mean-Value Theorem, we can get $\boldsymbol{\beta}^\ast$ lying on the segment joining $\boldsymbol{\beta}_{0S}$
and $\widehat{\boldsymbol{\beta}}_S$ such that
$$
\nabla \widetilde{L_n^P}(\widehat{\boldsymbol{\beta}}_S) = \nabla \widetilde{L_n^P}({\boldsymbol{\beta}}_{0S})
+ \nabla^2 \widetilde{L_n^P}({\boldsymbol{\beta}}^\ast)(\widehat{\boldsymbol{\beta}}_S - \boldsymbol{\beta}_{0S}).
$$
Therefore, denoting $\boldsymbol{D}=\left[\nabla^2 \widetilde{L_n^P}({\boldsymbol{\beta}}^\ast) 
- \nabla^2 \widetilde{L_n^P}({\boldsymbol{\beta}}_{0S})\right](\widehat{\boldsymbol{\beta}}_S - \boldsymbol{\beta}_{0S})$, we get
$$
\nabla^2 \widetilde{L_n^P}({\boldsymbol{\beta}}_{0S})(\widehat{\boldsymbol{\beta}}_S - \boldsymbol{\beta}_{0S})
+\boldsymbol{D} = -P_n'(|\widehat{\boldsymbol{\beta}}_S|)\circ \mbox{sgn}(\widehat{\boldsymbol{\beta}}_S) -  
\nabla \widetilde{L_n^P}({\boldsymbol{\beta}}_{0S}).
$$
Now, take any unit vector $\boldsymbol{\alpha}\in \mathbb{R}^s$. 
Then, since $\nabla^2 \widetilde{L_n^P}({\boldsymbol{\beta}}_{0S}) = \boldsymbol{\Sigma} + o_P(1)$ by definition, 
using the consistency of $\widehat{\boldsymbol{\beta}}_S$ we have from the above equation that
\begin{eqnarray}
\sqrt{n}\boldsymbol\alpha^t \boldsymbol{\Gamma}^{-1/2}\boldsymbol{\Sigma} (\widehat{\boldsymbol{\beta}}_S-\boldsymbol\beta_{0S}) 
= - \sqrt{n}\boldsymbol\alpha^t \boldsymbol{\Gamma}^{-1/2}\nabla \widetilde{L_n^P}({\boldsymbol{\beta}}_{0S})
- \sqrt{n}\boldsymbol\alpha^t \boldsymbol{\Gamma}^{-1/2}\left[P_n'(|\widehat{\boldsymbol{\beta}}_S|)\circ \mbox{sgn}(\widehat{\boldsymbol{\beta}}_S)
+\boldsymbol{D}\right].~~
\label{EQ:c1}
\end{eqnarray}

To tackle the first term in (\ref{EQ:c1}), we recall that 
$\nabla \widetilde{L_n^P}(\boldsymbol{\beta}_{0S}) = 2 \boldsymbol{A}_n(\boldsymbol{\beta}_{0S})\boldsymbol{J}(\boldsymbol{\beta}_0)
\boldsymbol{B}_n$, where
the random component 
$\boldsymbol{B}_n = \left[\frac{1}{n}\boldsymbol{\Pi}_S\widetilde{\boldsymbol{V}}_z^{-1}(\boldsymbol{y} - \boldsymbol{X}_S\boldsymbol{\beta}_S)\right]$
is normally distributed with 
$$
\mbox{Var}(\sqrt{n}\boldsymbol{B}) = \frac{\sigma^2}{n}\boldsymbol{\Pi}_S\widetilde{\boldsymbol{V}}_z^{-1}
\boldsymbol{V}(\boldsymbol{\theta},\sigma^2)\widetilde{\boldsymbol{V}}_z^{-1}\boldsymbol{\Pi}_S
\rightarrow \boldsymbol{\Upsilon}, ~~\mbox{ as }~n\rightarrow\infty.
$$
So, by the central limit theorem, for any unit vector $\widetilde{\boldsymbol{\alpha}}\in \mathbb{R}^{2s}$, 
$$
\sqrt{n}\widetilde{\boldsymbol\alpha}^t \boldsymbol{\Upsilon}^{-1/2}\boldsymbol{B}_n \mathop\rightarrow^\mathcal{D}N(0,1).
$$
Further, by definition $||\boldsymbol{A}_n(\boldsymbol{\beta}_0) - \boldsymbol{A}||=o_P(1)$. Hence, by Slutsky's theorem, we have
\begin{eqnarray}
\sqrt{n}\boldsymbol\alpha^t \boldsymbol{\Gamma}^{-1/2}\nabla \widetilde{L_n^P}({\boldsymbol{\beta}}_{0S}) \mathop\rightarrow^\mathcal{D}N(0,1).
\label{EQ:c2}
\end{eqnarray}

Next, for the second term in (\ref{EQ:c1}), we apply Lemma C.2 of Fan and Liao (2014) to get, under Assumption (P), 
$$
\left|\left|P_n'(|\widehat{\boldsymbol{\beta}}_S|)\circ \mbox{sgn}(\widehat{\boldsymbol{\beta}}_S)\right|\right|
=O_P\left(\max_{||\boldsymbol{\beta}_S-\boldsymbol{\beta}_{0S}||\leq d_n/4}\zeta(\boldsymbol{\beta}) \sqrt{\frac{s\log p}{n}} 
+ \sqrt{s}P_{n,\lambda}'(d_n)\right).
$$
Also, by Assumptions (I4)--(I5), we have $\lambda_{\min}(\boldsymbol{\Gamma}^{-1/2}) = O_P(1)$. Hence, applying Assumptions (A1)--(A2), we get

\begin{eqnarray}
&&\lambda_{\min}(\sqrt{n}\boldsymbol{\Gamma}^{-1/2})\left|\left|P_n'(|\widehat{\boldsymbol{\beta}}_S|)\circ \mbox{sgn}(\widehat{\boldsymbol{\beta}}_S)\right|\right|
\nonumber\\
&&~~\leq O_P(\sqrt{n})O_P\left(\max_{||\boldsymbol{\beta}_S-\boldsymbol{\beta}_{0S}||\leq \frac{d_n}{4}}\zeta(\boldsymbol{\beta}) \sqrt{\frac{s\log p}{n}} 
+ \sqrt{s}P_{n,\lambda}'(d_n)\right) = o_P(1).
\nonumber
\end{eqnarray}
Further, by continuity of $\nabla^2\widetilde{L_n^P}(\boldsymbol{\beta}_S)$, one can easily show that 
$$
\left|\left|\nabla^2 \widetilde{L_n^P}({\boldsymbol{\beta}}^\ast) 
- \nabla^2 \widetilde{L_n^P}({\boldsymbol{\beta}}_{0S})\right|\right|=o_P\left(\frac{1}{\sqrt{s\log p}}\right).
$$
Also, we have $||\widehat{\boldsymbol{\beta}}_S - \boldsymbol{\beta}_{0S}||=O_P\left(\sqrt{\frac{s\log p}{n}} + \sqrt{s}P_{n,\lambda}'(d_n)\right)$.
Then, combining the above equations with Assumption (A1), we have 
$||\boldsymbol{D}||=o_P(n^{-1/2})$. Hence, we get 
\begin{eqnarray}
\sqrt{n}\boldsymbol\alpha^t \boldsymbol{\Gamma}^{-1/2}\left[P_n'(|\widehat{\boldsymbol{\beta}}_S|)\circ \mbox{sgn}(\widehat{\boldsymbol{\beta}}_S)
+\boldsymbol{D}\right]=o_P(1).
\label{EQ:c3}
\end{eqnarray}

\noindent
Therefore, using (\ref{EQ:c2}) and (\ref{EQ:c3}) in (\ref{EQ:c1}) with the help of Slutsky's theorem, we get the desired asymptotic normality result
completing the proof of the theorem.
\hfill{$\square$}




	%

\end{document}